\documentclass[aip,graphicx, reprint]{revtex4-1}
\usepackage{dcolumn}
\usepackage{bm}

\usepackage[utf8]{inputenc}
\usepackage[T1]{fontenc}
\usepackage{mathptmx}
\usepackage{etoolbox}
\usepackage{mathtools}
\usepackage{xcolor}
\usepackage[normalem]{ulem}
\usepackage{verbatim}
\usepackage{graphicx}
\newcommand*{\kb}{k_{\rm{B}}}
\newcommand{\intext}[1]{\textcolor{black}{#1}}
\newcommand{\outtext}[1]{}

\begin{document}


\title{Electronic measurements of entropy in meso- and nanoscale systems} 




\author{Eugenia Pyurbeeva}
\email{e.d.pyurbeeva@qmul.ac.uk}
\affiliation {\small \textit School of Physics and Astronomy, Queen Mary University of London, Mile End Road, London E1 4NS, UK}
\author{Jan A. Mol}
\affiliation {\small \textit School of Physics and Astronomy, Queen Mary University of London, Mile End Road, London E1 4NS, UK}
\author{Pascal Gehring}
\email{pascal.gehring@uclouvain.be}
\affiliation {\small \textit IMCN/NAPS, Université Catholique de Louvain (UCLouvain), 1348 Louvain-la-Neuve, Belgium}

%

\date{\today}

\begin{abstract}
Entropy is one of the most fundamental quantities in physics. For systems with few degrees of freedom, the value of entropy provides a powerful insight into its microscopic dynamics, such as the number, degeneracy and relative energies of electronic states, the value of spin, degree of localisation and entanglement, and the emergence of exotic states such as non-Abelian anyons. As the size of a system decreases, the conventional methods for measuring entropy, based on heat capacity, quickly become infeasible due to the requirement of increasingly accurate measurements of heat. Several methods to directly measure entropy of mesoscopic quantum systems have recently been developed. These methods use electronic measurements of charge, conductance and thermocurrent, rather than heat, and have been successfully applied to a wide range of systems, from quantum dots and molecules, to quantum Hall states and twisted bilayer graphene. In this Review, we provide an overview of electronic direct entropy measurement methods, discuss their theoretical background, compare their ranges of applicability and look into the directions of their future extensions and applications. 

\end{abstract}


\maketitle 
\tableofcontents

\section{Introduction}
\label{introduction-section}
The connection between macroscopic observables and microscopic dynamics was first made by Maxwell when he attributed part of the heat capacity $C$ of a gas to the rotational degrees of freedom of its molecules\cite{Maxwell1860}. The name \emph{entropy} was coined by Clausius in 1865 to describe the non-usable energy increase in a steam engine's exhaust as a function of its temperature $T$ through the relation $dS=CdT/T$. Boltzmann and Gibbs later gave entropy its statistical basis that connects the observable averaged state functions to the microscopic dynamics of a system described by the number of microstates $\Omega$ and their probabilities, epitomised by Boltzmann's constant $\kb$ in his definition of entropy $S = \kb \ln \Omega$. Since then, entropy measurements based on heat capacity and temperature have been used to probe the microscopic structure of bulk materials, such as the disorder in alloys \cite{Bragg1934, Bragg1935, Nix1938}, phase transitions in molecular crystals \cite{Guthrie1961} and the magnetic arrangements of spin-ice \cite{Ramirez1999}.

The development of nanofabrication, in particular molecular beam epitaxy and electron beam lithography, has made it possible to confine electrons in one or more dimensions. This has enabled the measurement of heat capacity and entropy in mesoscopic systems such as two-dimensional (2D) electron gasses in GaAs quantum well structures \cite{Gornik1985, Wang1988, Bayot1996} and of fractional quantum Hall systems \cite{Schulze-Wischeler2007, Schmidt2017}. These experiments present significant difficulties as the contribution to the heat capacity of the confined electronic states is generally far smaller than that of the vibrations, or phonons, in the substrate material. To overcome this challenge, the electronic contribution to the heat capacity was increased by increasing the number of quantum wells \cite{Gornik1985,Bayot1996, Wang1988} and by exploiting the difference in thermalisation time of the electronic and phononic systems \cite{Schulze-Wischeler2007, Schmidt2017}. Despite the ingenuity of these solutions, further down-scaling of entropy measurements based on heat capacity to few-electron nanodevices is practically impossible as the small numbers of electrons and electronic states involved require the measurement of increasingly small amounts of heat and heat capacities. For comparison, the maximum heat capacity of a single electronic spin is of the order of $10^{-24}$J/K, while the minimal values of heat capacity per particle experimentally measured in \cite{Schmidt2017} are three orders of magnitude greater.  

Yet, it is precisely \outtext{for a system}\intext{in systems} containing only a few particles that entropy measurements can reveal the most about \outtext{its} microscopic dynamics. For example, the von Neumann entropy of the entangled state $\left(\vert00\rangle + \vert11\rangle\right)/\sqrt{2}$ is $k_\text{B}\ln2$, while the entropy of the product state $\left(\vert00\rangle + \vert01\rangle+\vert10\rangle+\vert11\rangle\right)/2$ is zero \cite{Potts, Kosloff2013a}. There is therefore a strong motivation to measure the entropy of few-electron quantum systems that has led to the development of alternative ways that do not rely on detecting the prohibitively small flow of heat. Electronic entropy measurements have recently emerged to address this challenge. Unlike heat, charge is a conserved quantity, which makes it much easier to measure. Charge sensitivities down to level of $10^{-6} \times $e/$\sqrt{\text{Hz}}$ can be achieved in few-electron devices, and charge currents as small as $10^4$ e/s can be detected. Both these measures, of charge state and current, can be used to determine the entropy of a system. Electronic entropy measurements are thus ideally suited to probe microscopic dynamics of few-electron quantum systems that are weakly coupled to their environment.

This Review will summarize the thermodynamic theory underpinning both charge- and current-based entropy measurements of mesoscopic quantum systems, and discuss their application to several material systems and devices, including \outtext{spin-ices and spin-frustrated systems\cite{Ramirez1999},} twisted bilayer graphene \cite{Rozen2021, Saito2021}, Kondo-systems\cite{Han2021}, quantum Hall states, and quantum dots and quantum dot systems \cite{Hartman2018, Kleeorin2019} as well as single-molecule devices \cite{Gehring2021, Pyurbeeva2021}. In all these systems electronic entropy measurements may serve a dual purpose. On the one hand, electronic entropy measurements provide insight into the internal quantum structure of matter, and might reveal exotic quasi-particles such as Majorana fermions. On the other hand, knowledge can be gained about the operation of an entire new class of nanoscale thermodynamic systems that include single-electron heat engines that operate close to the Curzon-Alhborn limit \cite{Josefssonb}, stochastic Maxwell demons \cite{Koski2015, Sanchez2019, Sanchez2019a} and Szilard engines \cite{Kim2011}, and clocks \cite{Erker2017, Pearson2021}.

\outtext{\textbf{Structure layout}. This Review is structured as follows: first, in Section \ref{continuous-section} we discuss the methods for measuring entropy through non-heat parameters in bulk materials, where charge, or particle number can be considered to be a continuous parameter. Then, in Section \ref{quantum-section} we look into the change in the approach in the case of quantised charge, the methods proposed for this case and their limitations. Section \ref{experiment-section} is devoted to the performed experiments in entropy measurement and potential systems that would benefit from entropic research. Finally, in Section \ref{final-section}, we conclude with a discussion of the limitations of electronic entropy measurements and the directions in which we see them further developed. }
\intext{In Section II of this review, we will discuss different approaches to determine entropy of bulk materials that do not rely on heat measurements. In Section III we will extend this discussion to systems with a quantised charge, and in Section IV we will discuss how these entropy measurements can be used to reveal microscopic dynamics of quantum systems under investigation. Finally, we will provide an outlook on the future of nanoscale thermodynamics and entropy measurements of quantum systems.}

\section{Alternative entropy measurements in continuous-charge nanodevices}
\label{continuous-section}
\begin{figure}[!ht]
\includegraphics[width=\linewidth]{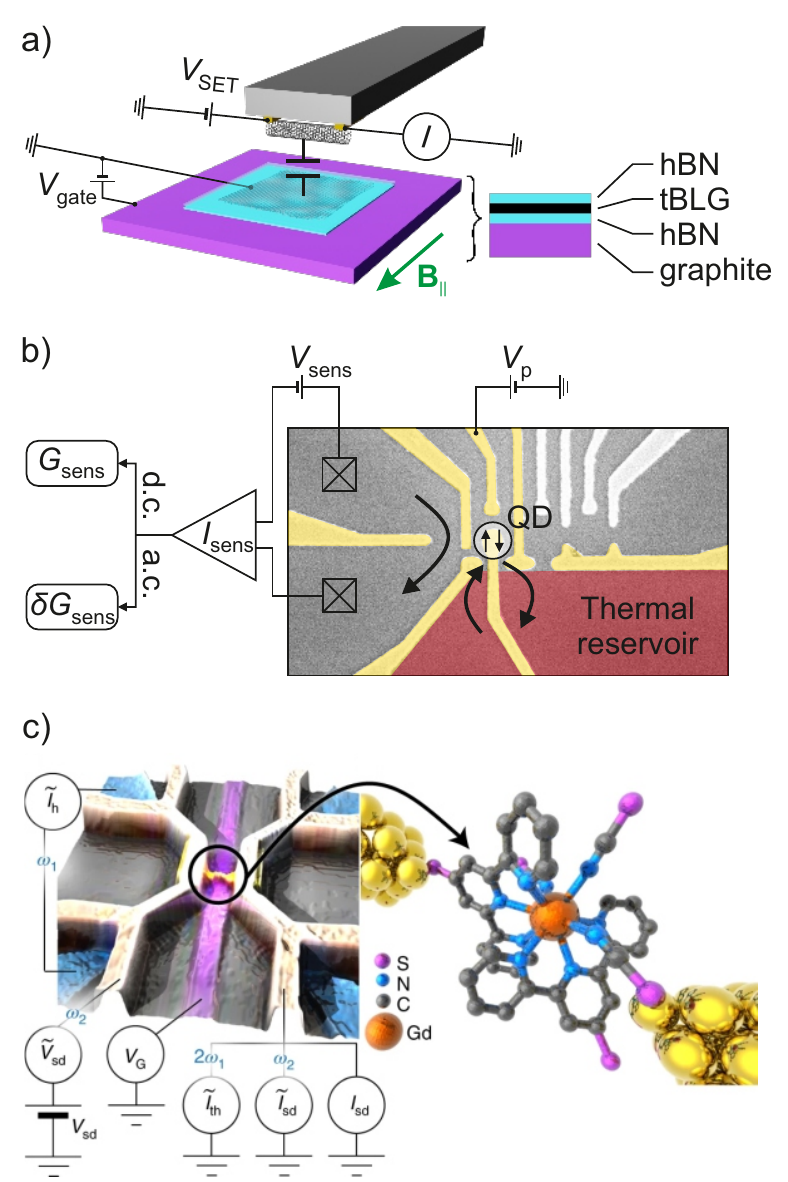}
\caption{\label{fig1} Experimental setups for measuring entropy in nanoscale systems. a) A nanotube-based single electron transistor (SET) measuring the charge density and compressibility in a continuous-charge system (twisted bi-layer graphene, tBLG) under an external magnetic field $B_{||}$\cite{Rozen2021}. The graphene is encapsulated from both sides by insulating hexagonal boron nitride (hBN) layers (blue). The graphite back gate (purple) is used for setting the chemical potential of the graphene. b) A charge-state based entropy measurement setup in a GaAs quantum dot\cite{Hartman2018}. \intext{The electrodes coloured in yellow electrostatically define the circuit in the 2D electron gas. The quantum dot (QD) exchanges electrons with a thermal electron reservoir (red). The quantum point contact on the left is used to measure the charge ($n$ (d.c.) and $\Delta n$ (a.c.), see Figure \ref{fig2}) of the quantum dot. This charge can be tuned by the voltage $V_\mathrm{p}$.}\outtext{The gold electrodes define the main quantum dot (red), a quantum point contact (QPC) of the left side of the device is used to control the temperature of the electron gas, while the QPC on the right is used as a charge sensor.} c) A transport entropy measurement setup in a single-molecule device\cite{Gehring2021}. A molecule bridges the gap formed by electromigration of gold contacts. Each side of the device has a heater (blue), while the back-gate (purple) is used to control the energy levels of the molecule. Simultaneous measurement of current, conductance and thermocurrent is possible due to the double demodulation technique -- the components of current through the molecule are taken at zero frequency, the oscillation frequency of the bias voltage and double the frequency of the heater bias.}
\end{figure}

\outtext{As was argued in the Introduction, the classical thermodynamic approach to measuring entropy is not suitable when the system size reaches mesoscopic or nanoscopic dimensions. Therefore, when the size of the system decreases, other parameters could be measured from which the entropy can be determined, and that are easier to access in experiments than heat flows.}
\intext{Since conventional approaches to entropy measurements are not easily applicable to mesoscopic systems, there has been great effort in developing approaches that rely on observables that are experimentally accessible in these systems, with electric measurements being the prime candidate. The thermoelectric effect, which connects heat (and therefore entropy) and charge transport has been proposed as an entropic probe for non-Abelian states\cite{Yang2009}. Indeed, for a free electron gas, the low-temperature Seebeck coefficient is directly proportional to the entropy per unit charge\cite{Chaikin1976,Goupil2016}. While this relation holds true in a wide range of bulk materials\cite{Behnia2004, Zlatic2007}, it breaks down, for example, in highly anisotropic transport\cite{Mravlje2016} and should thus be used with caution in mesoscopic systems.}

\outtext{The first example of utilising the connection between entropy and electric measurements comes from thermoelectrics -- a field which by definition links heat (entropy) and charge transport. It has long been known that the Seebeck coefficient is a measure for the entropy per unit charge carrier \cite{Chaikin1976}.}  

\outtext{For a free electron gas the low-temperature Seebeck coefficient (often referred to as thermopower) is directly proportional to the entropy of a unit charge carrier \cite{Goupil2016}. This holds true in a wide range of bulk solid materials \cite{Behnia2004, Zlatic2007}, where the Seebeck coefficient can be used to measure entropy, however does not apply to all materials. For instance, the relation between the Seebeck coefficient and entropy falls apart in systems with highly anisotropic transport\cite{Mravlje2016}. Despite this, thermopower has been proposed as an entropic probe for non-Abelian states\cite{Yang2009}.}

\outtext{Maxwell relations are a textbook source of thermodynamically justified connections between the derivatives of various quantities. It was first proposed by Cooper and Stern\cite{Cooper2009} to employ them for the observation of entropic signatures of non-Abelian quantum Hall states, as an alternative to the interferometry methods proposed at the time \cite{Sarma2005, Stern2006, Bonderson2006, Feldman2007}. The work\cite{Cooper2009} uses two Maxwell relations:}
\intext{Going beyond the thermoelectric relation between heat and charge transport, Maxwell relations can be used to connect the experimentally controllable magnetic field, $B$, and temperature, $T$, and the observable magnetisation, $m$, electron density, $n$, and chemical potential, $\mu$, of a quantum Hall system to its entropy per unit area, $s$\cite{Cooper2009}:}
\begin{equation}
\label{maxwell-1}
    \left( \frac{\partial m}{\partial T}\right)_B= \left( \frac{\partial s}{\partial B} \right)_T
\end{equation}
\begin{equation}
\label{maxwell-2}
    \left( \frac{\partial \mu}{\partial T}\right)_n= -\left( \frac{\partial s}{\partial n} \right)_T
\end{equation}
\outtext{where $B$ and $T$ are the magnetic field and temperature -- free parameters for the experimentalist, $m$, $n$ and $\mu$ are the magnetisation density, the electron surface density and the chemical potential -- quantities that can be measured, and $s$ is the entropy per unit area, which can reveal the non-Abelian degrees of freedom.} 

\outtext{At the same time, nothing in equations \ref{maxwell-1} and \ref{maxwell-2} is specific to the quantum Hall effect. Indeed, the authors\cite{Cooper2009} suggest the application of the same Maxwell relation method to any electronic system with temperature- and field-dependent magnetisation, such as spin-ordering transitions in low density electron gases in the Wigner crystal or Luttinger liquid states. Additionally equations similar to \ref{maxwell-1} and \ref{maxwell-2} can be written for many parameters, to explore systems with different effects, such as pressure and volume or electric field and dipole moment.}
\intext{This approach was initially proposed as an alternative to interferometry methods to detect signatures of non-Abelian quantum Hall states\cite{Sarma2005, Stern2006, Bonderson2006, Feldman2007}, however it can equally be applied to any electronic system with temperature- and field-dependent magnetisation, such as spin-ordering transitions in low density electron gases in Wigner crystal or Luttinger liquid states. Maxwell relations have also been used to study magic-angle twisted bi-layer graphene to reveal a Pomeranchuk-type effect\cite{Rozen2021,Saito2021} (see Section \ref{graphene-section}) in a setup as shown in Figure 1a.}

\intext{While the thermoelectric and Maxwell relation approaches that we have described thus far are generally system-independent they rely on the implicit assumption of large particle numbers such that any change in state variables such as entropy and magnetisation is linear with the change in particle number\cite{Chaikin1976, Zlatic2007, Yang2009, Mravlje2016}. This assumption does not necessarily hold for meso- and nanoscale systems with small numbers of particles. In the remainder of this Review we will discuss the conceptual and experimental differences between entropy measurements in what we term continuous-charge devices and quantised-charge devices.}

\section{Quantised-charge nanodevices}
\label{quantum-section}

\begin{figure*}[!ht]
\includegraphics[width = \linewidth]{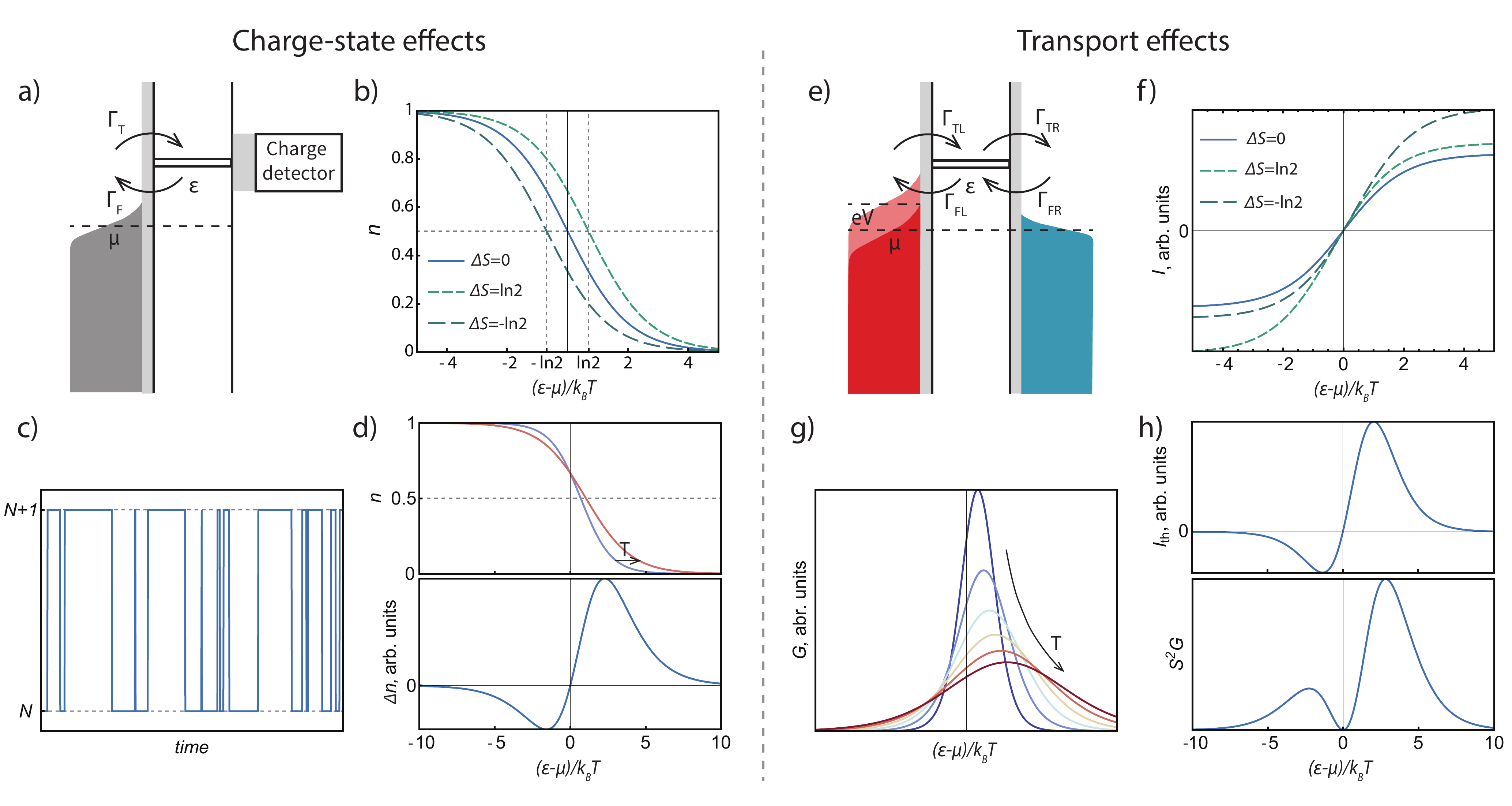}
\caption{\label{fig2} Entropic effects in charge state and charge transport measurements in quantised-state nanodevices.  a) A schematic of a charge state measurement setup. An energy level coupled to a thermal bath, with the charge state of the level independently determined. b) The mean excess population $n$ of the energy level in the case of the non-degenerate and two-fold degenerate energy level ($\Delta S=0$ or $\Delta S=\pm \ln2$). c) A model of experimental time-resolved measurement of the charge state of the device. Information of entropy (or degeneracy) can be extracted from the proportion of time the device spends in each of the charge states\cite{Hofmann2016}. d) A model of the experiment performed in \cite{Hartman2018} -- mean population is measured as a function of gate voltage (energy level), where an overall shift of the distribution is observed on top of its spreading. The exact value of the entropy difference can be found from fitting the difference between the populations at the two temperatures. e) A schematic of a charge transport measurement setup. The energy level is coupled to two thermal baths, with a potential and/or a temperature difference across. The current through the system is measured. f) The current through the nanodevice for a non-degenerate and two-fold degenerate energy level ($\Delta S=0$ or $\Delta S=\pm \ln2$). g) A model of the experimental thermal shift of the conductance peak of a device, as observed in \cite{Harzheim2020, Gehring2021}. h) A model of the thermocurrent and power factor measurement of a quantised-charge nanodevice. The value of the entropy difference can be extracted from the asymmetry. }
\end{figure*}

\subsection{The effects of charge quantisation}
\intext{We define quantised-charge devices as those meso- and nanoscale systems for which partial derivatives with respect to particle number $N$ are undefined.}\outtext{Quantised-charge nanodevices are those where the derivatives by particle number $N$ can not be defined.} \outtext{One must}\intext{It is important to} note that \outtext{a quantised-charge nanodevice does}\intext{these devices do} not necessarily contain a smaller total number of particles than a continuous-charge \outtext{one}\intext{devices, but rather that they have a non-linear energy-particle relation and typically contain a smaller number of available particle number states.}\outtext{. The difference lies in the number of available particle number states and the dependence of energy on them rather than the overall particle number. A well-defined derivative requires the existence of a linear approximation of energy as a function of particle number for several values of the latter.

} The most common example of \outtext{a system without a derivative by particle number are nanodevices in the state of}\intext{quantised-charge devices are meso- or nanoscale systems that exhibit} Coulomb blockade\cite{Nazarov2009}, \outtext{where}\intext{meaning} the electrostatic \intext{Coulomb }repulsion is the dominant energy scale and only \intext{a small number of} charge states (values of particle number in the system) are energetically \outtext{allowed}\intext{accessible}. \outtext{Typical examples for such systems are}\intext{Coulomb blockade is regularly observed in devices based on lithographically-defined }quantum dots\cite{Hanson2007, Thijssen2008} (see Figure \ref{fig1}b)\outtext{ and molecular devices in the weak coupling transport regime}\intext{, and in single-atom\cite{Fuechsle2012} and single-molecule devices}\cite{Osorio2008, Brooke2015} (see Figure \ref{fig1}c). 

\outtext{In such devices,}\intext{The change of} the electrostatic energy \outtext{change with the increase of particle number }depends on the particle number itself \intext{as}
\begin{equation}
    E_{el}(N+1)-E_{el}(N)=2e^2/C\left( N+\frac{1}{2}-N_0 \right)
\end{equation}
where $E_{el}$ is the electrostatic energy \intext{of the system}, \outtext{$E_C$ is the charging energy with $E_C=e^2/C$, }$C$ is \outtext{the system's}\intext{its} capacitance, and $N_0$ is the \outtext{induced }\intext{background }charge\intext{.}\outtext{, which is a function of gate voltage.} \intext{ Therefore, a partial derivative of energy with respect to particle number cannot be defined for systems in which the electrostatic energy dominates, even though they may contain a large number of particles, $N_0$.}
\outtext{Thus, both a single molecule in the sequential tunnelling regime (Fig \ref{fig1}c) and a GaAs quantum dot with a characteristic size of hundreds of nanometer (Fig \ref{fig1}b) are quantised-charge nanodevices in the sense that derivatives by particle number cannot be defined in them, due to the significant electrostatic contribution to the energy, despite having vastly different overall numbers of electrons. A GaAs quantum dot, while being a quantised-charge device, might have more electrons in total than a given sample of magic-angle graphene in the continuous-charge regime (Fig \ref{fig1}a). Most frequently, Coulomb-blocked devices are considered in the case where only two charge states are energetically available: $N$ and $N+1$. In this case no linearisation of any parameter by particle number can be justified, if $N$ and $N+1$ are the only two available charge states:}\intext{Moreover, in Coulomb-blocked devices where only two charge states are energetically accessible and $\Delta N$=1, no linearisation of any state variable $X$ can be justified, since}
\begin{equation}
\label{can't-do}
    \left(\frac{\partial X}{\partial N}\right)_{Y}\neq \frac{X(N+1)-X(N)}{\Delta N}=X(N+1)-X(N).
\end{equation}
\outtext{where $X$ is some variable.} Additionally, the partial derivative at constant $Y$ \intext{cannot} be defined, as each charge state is associated with its own value of $Y$. This means that the approach \outtext{outlined in section \ref{continuous-section}}\intext{developed for continuous-charge devices} cannot be applied directly \intext{to quantised-charge devices}.

\subsection{Thermodynamic parameters of a quantised-charge system } 
\label{parameter-section}
One of the main strengths of thermodynamics is system-independence. \outtext{The}\intext{A} thermodynamic treatment of quantised-charge devices\outtext{can allow}\intext{, analogous to the Maxwell relations for continuous-charge devices, allows} for the development of entropy measurement methods that  \outtext{apply not only to devices with a single degenerate energy level, discussed in Section \ref{degeneracy-section}, but to those with more}\intext{describe the} general \intext{quantum} dynamics \intext{of such systems}.\outtext{It can be done in a way  analogous to the continuous-charge regime, discussed in Section \ref{continuous-section}: starting from thermodynamic principles and using Maxwell relations. However, to} \intext{To} do this\outtext{ in quantised-charge systems, one has to}\intext{, we will first} specify the parameters \intext{(state variables)} used in the thermodynamic description of \outtext{a }small quantised-charge system\intext{s}, namely the temperature $T$, chemical potential $\mu$, particle number $N$, and entropy $S$. 
\outtext{Below, we treat them separately.}

\textbf{Temperature}. 
While a quantised-charge system, or its electronic state dynamics may or may not be sufficiently large to have its own definable temperature, the Maxwell equation approach taken in section \ref{continuous-section} requires equilibrium conditions. Following this, we may take the temperature of the nanoscale system to be equal to the well-defined temperature of the bath \intext{with which } it exchanges electrons \outtext{with} \cite{Blundell, Ford2013, Potts}. In the case of more than one bath being present, equilibrium is only possible when the temperature differences between the baths are infinitesimally small\outtext{, and the temperature of the system can be defined to a small window}.

\textbf{Chemical potential}.
\outtext{It is customary in the quantum dot community to call}\intext{In quantum dot literature,} the energy difference between the charge states $\epsilon = U(N+1)-U(N)$ \intext{is often referred to as} the chemical potential \intext{$\mu$}\cite{Hanson2007}. However, \outtext{it}\intext{this} does not agree with the canonical definition of the chemical potential \intext{$\mu=(\partial U/\partial N)_S$ due to reasons outlined after equation \ref{can't-do}.} 
\outtext{not only due to the absence of the derivative, but also as the effect that is studied in the dependence of entropy on charge state, thus changing the charge state at constant entropy is impossible. 

}Instead, for quantised-charge systems, one should take the approach similar to that of temperature \cite{Pyurbeeva2021, Child2021a} -- if a system is in equilibrium with a bath regarding particle exchange, its chemical potential is equal to that of the bath \cite{Blundell, Ford2013}. And if several baths are present, for quasi-equilibrium conditions the difference in chemical potentials must be infinitesimally small. For all realistic temperatures \intext{at which} nanoelectronic measurements are performed, the electron gas \intext{that forms the bath} is highly degenerate and the chemical potential \intext{of the bath and the system} \outtext{is}\intext{are} to a high degree of accuracy equal to the Fermi energy, $\mu=E_F$. 

\textbf{Particle number}.
\outtext{In the Coulomb-blocked state we consider, the system only allows two particle numbers, $N$ and $N+1$, and in the presence of particle exchange with the bath, the particle number is not only a function of external parameters (temperature, chemical potential, magnetic field, etc) as it is in the continuous case (section \ref{continuous}), but also of time. To simultaneously remove this dependence and large fluctuations $\Delta N=1$ as well as the discontinuity, the mean number of particles can be used as a variable:}
\intext{When we consider a quantised-charge device that exhibits Coulomb blockade, the particle number constantly fluctuates between $N$ and $N+1$ with occupation probabilities $p_N$ and $p_{N+1}$ that depend on external parameters (temperature, chemical potential, magnetic field, etc.) similarly to $N$ in continuous-charge devices. The problem of temporal charge fluctuations and the discontinuity of state variables can be overcome by using the mean particle number}
\begin{equation}
    \bar{N}=p_N N+ p_{N+1} (N+1)=N+n,
\end{equation}
\outtext{Here, one can define}\intext{where $n=p_{N+1}$ is defined as} the mean excess number of electrons \outtext{$n=p_{N+1}$}, which \outtext{changes}\intext{varies} between 0 and 1\outtext{, as $N$ remains constant in our consideration}. 

\textbf{Entropy}.
In line with the \intext{time-independent} mean particle number \outtext{ used in the thermodynamic description of quantised-charge nanodevices, which does not depend on time, one has to}\intext{we can} define \intext{the mean} entropy \outtext{in a similar way, }for a state with a given \outtext{value of}\intext{mean excess number of electrons} $n$. The entropy of a system in each of the charge states, $S_N$ or $S_{N+1}$, is easily defined, however in the case of fluctuating charge, an additional term is needed to account for the uncertainty of the charge state itself \outtext{. It can be shown that entropy in that case is equal to}\intext{ and the mean entropy of the system is given by}\cite{Pyurbeeva2021}
\begin{equation}
\label{entropy}
    \bar{S}=S_0+nS_{N+1}+(1-n)S_{N},
\end{equation}
where $S_0$ is the ``coarse-grained'' Gibbs entropy of the charge state uncertainty $S_0=-n \ln n - (1-n) \ln (1-n)$.

\subsection{Maxwell relations in quantised-charge systems}
Taking the considerations for the treatment of independent thermodynamic parameters outlined in Section \ref{parameter-section}, the standard Maxwell relation should be interpreted as 
\begin{equation}
\label{Maxwell}
    \left( \frac{\partial \mu}{\partial T} \right)_{\bar{N}}=-\left( \frac{\partial \bar{S}}{\partial \bar{N}} \right)_T
\end{equation}
where $\mu$ is the chemical potential of the bath, $\bar{N}$ is the mean time-averaged population of the system and $\bar{S}$ is the Gibbs entropy incorporating both charge state uncertainty and the entropy of \outtext{the}\intext{each} charge state\outtext{s}.\outtext{ 

The mean free energy of the system can be expressed as:
\begin{equation}
    \bar{F}=E(N)+\varepsilon  n -TS
\end{equation}
where $\varepsilon=E(N+1)-E(N)$ is the energy of the transport level \outtext{described in section \ref{degeneracy-section}}, which can be experimentally varied by the application of gate voltage, $V_G$: $\varepsilon=\varepsilon_0+\alpha V_G$, where $\alpha$ is the lever arm, quantifying the electrostatic coupling between the system and the gate electrode. Using this and the \intext{general} relation for the chemical potential $\mu = \left( \partial \bar{F} / \partial \bar{N}  \right)_T$, one arrives at:} \intext{ However, since it is the difference between the single-particle energy of the system and the chemical potential of the bath that is proportional to the externally controlled gate voltage applied to a quantised-charge device, $\epsilon-\mu\propto V_G$, it is more useful to derive a Maxwell relation for this difference rather than for the chemical potential. }

\intext{Such a relation, which we name the \emph{Microscopic Maxwell Relation}
can be derived by considering the mean free energy of the system, $\bar{F}=U(N)+\varepsilon  n -T\bar{S}$, and the general relation relation for the chemical potential $\mu = \left( \partial \bar{F} / \partial \bar{N}  \right)_T$, which yields}
\begin{equation}
\label{almost}
    \varepsilon-\mu=T \left( \frac{\partial \bar{S}} {\partial \bar{N}}\right)_T .
\end{equation}

\outtext{In a two-charge-state system all derivatives by $\bar{N}$ can be replaced by those by $n$, since $N$ is constant. Rewriting equation \ref{almost} in this way in conjunction with the two-state entropy expression (eq. \ref{entropy}), we arrive at the \emph{Microscopic Maxwell Relation}\cite{Pyurbeeva2021}:}\intext{Since $N$ is constant we can replace all derivatives with respect to $\bar{N}$ with derivatives with respect to $n$ and arrive at what we call the \emph{Microscopic Maxwell Relation}\cite{Pyurbeeva2021}} 
\begin{equation}
\label{maxwell_gen}
   \frac{\varepsilon-\mu}{T}=\kb \ln \frac{1-n}{n}+\Delta S.
\end{equation}\outtext{

This relation can be independently proven\cite{Pyurbeeva2021} starting from the Gibbs distribution. Another property of Equation \ref{maxwell_gen} is that the population of a system with an entropy difference $\Delta S$ between the charge states is:
\begin{equation}
    n=\frac{1}{1+e^{\frac{\varepsilon-\mu-T\Delta S}{\kb T}}},
\end{equation}
which is a Fermi-distribution shifted by $T \Delta S$.

A}\intext{In a} different approach, \outtext{one that can be called an}\intext{which we name the} \emph{Integral Maxwell Relation} \outtext{involves reformulating }the Maxwell relation (equation \ref{Maxwell}) \intext{can be reformulated in terms of the }\outtext{in the same sense of }time-averaged particle number $\bar{N}$ and the chemical potential of the bath as\cite{Sela2019, Child2021, Child2021a}
\begin{equation}
\label{maxwell_int}
    \Delta S_{\mu_1 \rightarrow \mu_2}=\int \limits_{\mu_1}^{\mu_2} \frac{d \bar{N}}{dT}d\mu.
\end{equation}
The benefit of Equation \ref{maxwell_int} is that it is completely equivalent to the Maxwell relation and therefore holds true for all systems. It has to be noted, however, that in it $\Delta S_{\mu_1 \rightarrow \mu_2}$ is the total entropy \intext {difference}, with no distinction between the entropy due to the charge state uncertainty and the entropy of the pure charge states. Thus, in order to find $\Delta S$ between the charge states, one has to shift $\mu$ (or, equivalently $\varepsilon$) from the point of a pure $N$ charge state to that a pure $N+1$ charge state: in the approximation of only two states being available, from $-\infty$ to $\infty$, but in reality between the two consecutive Coulomb-blocked regions. 

\subsection{Quantised-charge entropy measurement methods}
\begin{figure*}[!ht]
\includegraphics[width = \linewidth]{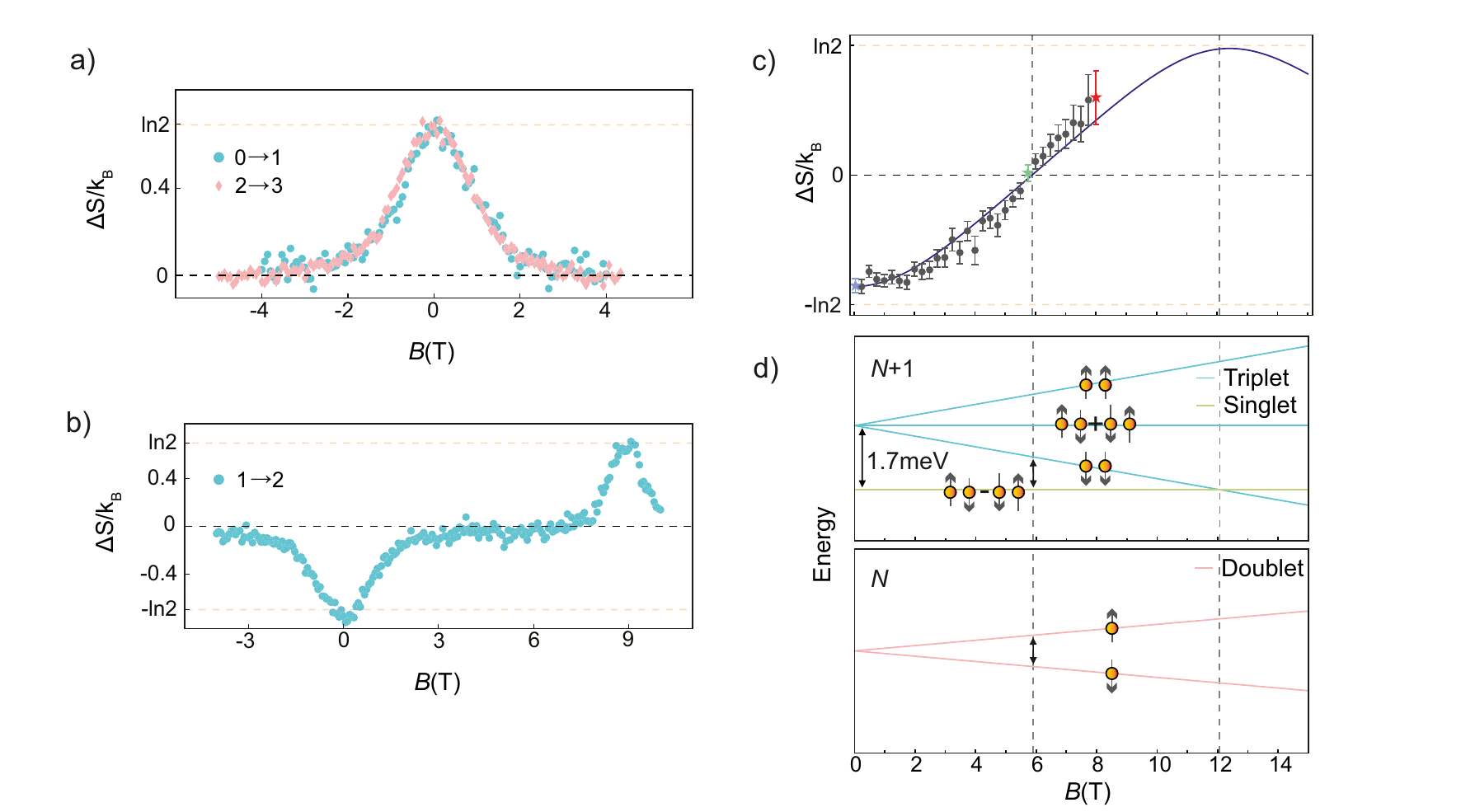}
\caption{\label{fig3} a) Experimental results for the entropy difference between the $N$ and $N+1$ charge states as a function of magnetic field \cite{Hartman2018}\outtext{for two values of even (left) and one of odd (right) $N$}. b) \intext{Experimental results for the entropy difference between the $N+1$ and $N+2$ charge states as a function of magnetic field \cite{Hartman2018}.\outtext{
The electronic entropy $s$ of twisted bilayer graphene \cite{Rozen2021} as a function of $\nu$, the moire lattice filling factor. \emph{Image modified from \cite{Rozen2021}.} } c) The entropy difference between two redox states of a single-radical molecule found through thermoelectric spectroscopy \cite{Pyurbeeva2021b}. The points\outtext{in the top panel} show the experimental data, while the solid line gives the fit to a doublet $\leftrightarrow$ singlet + triplet transition model \intext{(panel d)}. The model parameters, such as the singlet-triplet splitting can be extracted from the fit. \intext{d) The fitting model for the entropy measurements in Fig. \ref{fig3} c.}   }}
\end{figure*}

Entropy measurement protocols can be divided into two groups -- those based on charge state measurement and those based on charge transport. 

\textbf{Charge-state based measurements} The first \outtext{proposed}\intext{demonstration of an} electronic entropy measurement method\cite{Hartman2018} was \intext{based on} a charge-state technique\outtext{. It } \intext{that }employed the thermal shift of the charge degeneracy ($n=1/2$) point (see Fig \ref{fig2}b) as a measure for entropy difference between the charge states. A quantum point contact (QPC) was located close to a quantum dot in a GaAs lithographically-defined device and used as a mean charge sensor (Fig. \ref{fig1}b). The dependence of mean charge of the dot on the transport energy level was measured at two \outtext{close} temperatures, and the entropy difference was found by fitting the asymmetry of the difference curve (see Fig. \ref{fig2}d).

The energy \outtext{of the level}\intext{$\epsilon$} at which the probabilities of both charge states are equal, \intext{denoted} $\varepsilon_{1/2}$, is \outtext{related to temperature and the entropy difference as:}\intext{given by Equation 9 with $n=1/2$}
\begin{equation}
    \varepsilon_{1/2}-\mu=T \Delta S
\end{equation}
which gives the thermal shift
\begin{equation}
\label{thermal-shift}
    \frac{\partial \varepsilon_{1/2}}{\partial T}=\Delta S.
\end{equation}
\outtext{In \cite{Hartman2018}, Equation \ref{thermal-shift} has only been proved for a degenerate level, however the same follows from equation \ref{maxwell_gen} by substituting $n=1/2$. The method described in \cite{Hartman2018} has been applied to a quantum dot with a two-fold degenerate level, yielding $\Delta S =\ln 2$, and the same dot in a magnetic field (see Fig \ref{fig3}a).}

Building on the first charge-state based entropy measurement method \cite{Hartman2018}, the \outtext{theoretical} applicability of the entropy measurement method for the detection of Majorana zero modes has been \outtext{demonstrated}\intext{theoretically shown}\cite{Sela2019} (see Section \ref{Majorana-section}) \outtext{by moving on to}\intext{using} the integral Maxwell relation formulation. Further \intext{experimental} developments concerned the design of a bespoke device and an optimised measurement protocol \outtext{optimising} \intext{improving} thermal equilibration in the system\cite{Child2021}, and the measurement of the entropy as a function of coupling strength (lifetime of a charge state), that could be controlled by gates forming the quantum dot\cite{Child2021a}.

\textbf{Charge transport based measurements} The first charge transport entropy measurement protocol \outtext{involves}\intext{that was experimentally demonstrated used} simultaneous measurement\intext{s} of conductance and thermocurrent and \intext{inferred the entropy from} fitting the thermocurrent \outtext{as:}\intext{to the following equation\cite{Kleeorin2019}}
\begin{equation}
    L(\varepsilon, T)=C(T)L_{NI}(\mu+\Delta(T),T)+G(\mu, T)\frac{\Delta(T)}{T},
\end{equation}
where $L$ is the thermoelectric conductance $L=\partial I/\partial \Delta T$ of the device, $L_{NI}$ is the thermoelectric conductance of the non-interacting (non-degenerate) level, which is symmetric in $\varepsilon$, $G$ is the \intext{electrical} conductance and $\Delta(T)$ is a function of entropy such that $\partial \Delta(T) / \partial T =\Delta S /2$. 

\intext{The first transport-based entropy measurement approach was subsequently generalised using the observation that Equation 9 can be written to give an expression of the excess number of electrons as a shifted Fermi distribution, $n = f(\epsilon-T\Delta S)$\cite{Pyurbeeva2021}, which yields the following equations for the conductance
\begin{equation}
\label{current-fit-method}
    G\propto \frac{1}{T}\left[ 1-f(\varepsilon)\right] f(\varepsilon-T\Delta S),
\end{equation}
and thermoelectric conductance\cite{Pyurbeeva2021b}
\begin{equation}
\label{thermocurrent-fit-method}
    L\propto \frac{\varepsilon}{T^2}\left[ 1-f(\varepsilon)\right] f(\varepsilon-T\Delta S).
\end{equation}
From this it follows that the energy for which the conductance is maximum, $\epsilon_G$, is given by 
\begin{equation}
    \varepsilon_G-\mu=\frac{T\Delta S}{2},
\end{equation}
and that therefore the thermal shift of the $G(\epsilon)$ peak position (see Fig. \ref{fig2}g) is\cite{Gehring2021}
 \begin{equation}
         \frac{\partial \varepsilon_G}{\partial T}=\frac{\Delta S}{2}.
 \end{equation}
Moreover, the entropy difference between two charge states can be directly measured from the asymmetry of the thermoelectric conductance $L(\epsilon)$ (see Fig. \ref{fig2}h).}
    
\subsection{Limitations of entropy measurement methods}
\label{limitations-section}
Both realisations of the Maxwell relation in microscopic systems (equations \ref{maxwell_gen}, \ref{maxwell_int}) follow from the standard thermodynamic Maxwell relation and therefore apply to a wide range of systems. \outtext{The most obvious group are those that can be represented by a set of arbitrarily spaced energy levels.} However, some physical properties \outtext{of} \intext{manifesting in} nanodevices are usually not considered in standard ``macroscopic'' thermodynamics. The two most obvious ones are \outtext{vibrations} \intext{vibrational effects} and lifetime broadening. 

It has been experimentally shown that mechanical degrees of freedom significantly affect charge transport in weakly coupled molecular devices \cite{Seldenthuis08, Thomas2019}. This was described by introducing a new energy parameter -- the reconfiguration energy $\lambda$ -- which quantifies the energy that is lost to the phonon bath due to the relaxation preformed by the molecule after an additional electron is added \cite{Sowa2019}. This energy can be associated with an entropy change $\Delta S_{ph}=\lambda/T_{ph}$ of the phonon bath, where $T_{ph}$ is its temperature. \intext{However, this loss of energy to the bath means that} \outtext{This is not a function of the state of the system because} each of the charge states can \outtext{not} \intext{no longer} be attributed to a particular entropy \intext{value}. \outtext{ due to the interaction between electronic and mechanical degrees of freedom. and sources of entropy.}Therefore, systems with \outtext{vibrational effects} \intext{strong coupling between electronic and mechanical degrees of freedom} are not described by the existing theory. At the same time, they offer a potential avenue of further work into electronic characterisation of mechanical states in systems with \outtext{large} \intext{strong} electronic-mechanical coupling.

In the rate equation approach lifetime broadening is represented by a Lorentzian-shaped peak replacing the $\delta$-function for the energy of the transport level \cite{Harzheim2020}. From the quantum-mechanical point of view, a finite electron lifetime in the charge states means that the number and energy operators, $\hat{N}$ and $\hat{\varepsilon}$, no longer commute. This invalidates\outtext{ the substitution of $\varepsilon$ for $\mu$ in} \intext{any equation containing $\varepsilon$ and $n$ simultaneously, such as} equation \ref{maxwell_gen}. However the integral form of the Maxwell relation (equation \ref{maxwell_int}) appears to hold, \intext{as it only uses the chemical potential of the bath,} and the dependence of entropy on coupling strength (lifetime) has been experimentally measured \cite{Child2021a}. \intext{A fundamental explanation of the measured entropy remains an interesting theoretical problem.} \outtext{An interesting intuitive result\cite{Child2021a} is that at high coupling strength (small lifetime) the entropy of a two-fold-degenerate quantum dot approaches the value of $\ln3$, representing maximal uncertainty between ``spin up'', ``spin down'' and ``no extra electron'' states.}  

Charge-transport based entropy measurement methods possess more limitations than the charge-state ones. Unlike mean charge, any transport property -- conductance, thermocurrent, power factor, etc. -- is fundamentally a non-equilibrium quantity and therefore is not necessarily linked to \intext{the state function} entropy. \outtext{ as a state function.} Furthermore, the derivation of all existing charge transport entropy measurement methods assume small level splitting -- a requirement that all energy states of the system observe the same electron population.  This requirement was well-formulated in the SI to \cite{Kleeorin2019}:\emph{ ``one of the levels dominates the transport, or the levels are degenerate, or at high temperatures ($T\gg \delta \varepsilon$) or at low temperatures \outtext{($T\gg \delta \varepsilon$)} \intext{($T\ll \delta \varepsilon$)}''}. \intext{At the same time, a strength of transport-based entropy measurement methods is that they can be performed on devices that are otherwise widely studied, usually for thermoelectric research. In some cases \cite{Kim2014, Dorsch2021}, they can even be used as a post-factum analysis of existing experimental data.}

\section{Current applications}
\label{experiment-section}

\subsection{Detection of non-Abelian anyons} 
\label{Majorana-section}
\intext{One of the motivating factors behind the development of electronic entropy measurement methods \cite{Hartman2018, Kleeorin2019} was to detect non-Abelian anyons, or Majorana modes.} \intext{Theory predicts that }\outtext{Theoretical studies suggest that} Majorana modes\outtext{ –- fermions that are their own antiparticle –- bear} \intext{have} the potential to \outtext{revolutionize}\intext{speed up and increase precision in} quantum computing\intext{\cite{Kitaev2003,Lian2018}}: Majorana qubits \outtext{should} support non-Abelian statistics, braiding and fault tolerant operation.

\outtext{Particularly promising are propagating 1D Majorana modes  as they  enable ultra-fast quantum computation\cite{Lian2018}.}

\outtext{Although}Several \outtext{material}\intext{physical} systems have been predicted to host Majorana modes, but their \outtext{identification, and more importantly, their control}\intext{detection and control} is very challenging and \outtext{scope of heavy ongoing research efforts}\intext{has not yet been achieved in a way that could be taken as proof beyond reasonable doubt}. Most \outtext{experiments}\intext{attempts} focus on electrical fingerprinting\cite{Sarma2005, Stern2006, Bonderson2006, Feldman2007}, but a key drawback of this approach is that the minute conductance signatures of Majorana modes could be easily misinterpreted or confused with \outtext{non-Majorana}\intext{those originating from other} mechanisms, especially in disordered samples\cite{Holden2021, Kayyalha2020, Zhang2021, Sills2021}.  

However, it has been predicted \intext{that nonlocal Majorana zero mode (MZM) states can be detected through their entropic signatures} \outtext{that entropy measurements could provide a robust signature of nonlocal Majorana zero mode (MZM) states}\cite{Cooper2009, BenShach2013}: while a Majorana qubit -- a nonlocal two-level system formed by two MZMs -- has a trivial entropy of $\kb \ln 2$, a single MZM should possess an universial fractional entropy of $\frac{1}{2} \kb \ln 2$ \cite{Smirnov2015} \intext{-- which is difficult to explain through different means}.

\outtext{This could be measured by the temperature dependence of charge transitions using a local charge detector.\cite{Hartman2018, Sela2019}}\intext{The main challenge is creating a system in which the entropy of a single quasiparticle can be measured in a charge measurement setup\cite{Hartman2018, Sela2019}. The }\outtext{Special care has then to be taken to reduce the}finite coupling between two MZMs forming the Majorana qubit \intext{can be reduced}\outtext{by} either \intext{by} increasing their spatial separation or by controlling the tunneling phases of the MZMs\cite{Smirnov2021a} via, for example, local gates controlling the tunneling barriers between topological superconductors and quantum dots\cite{Gau2020a,Gau2020b}. \outtext{Such}\intext{Alternatively,} experiments can be simplified by increasing the temperature, which diminishes the dependence of \outtext{the} entropy on the MZM tunneling phases, and by involving multiple MZMs. In the latter case the sensitive adjustment of the tunneling phases becomes obsolete and an entropy of $\frac{n}{2}\kb \ln 2$, where $n$ is the number of MZMs, is expected.\cite{Smirnov2021b}

A single spin-1/2 impurity that is anti-ferromagnetically coupled to a single conduction electron channel \outtext{can give}\intext{gives} rise to \outtext{a} \textit{single channel Kondo} effect. Such Kondo effect is universal and its universality has recently been verified by thermocurrent spectroscopy\cite{Hsu2022}. At high temperatures the impurity is free and acts as an unpaired spin with entropy $\kb \ln 2$\outtext{)}, \intext{while} at low temperatures \outtext{the impurity}\intext{it} is screened by \outtext{the} conduction electrons (a "Kondo cloud"),\outtext{ and a many-body-singlet forms that has} \intext{forming a many-body-singlet with} zero residual entropy. However, \outtext{if there are}\intext{in the case of} multiple conduction electron channels the impurity can be "overscreened" and a finite \intext{(non-zero)} residual entropy is expected. \outtext{Depending on the exact number of channels,}\intext{It has been predicted that} such systems can host Majorana fermions \outtext{(for two channels)} with a residual impurity entropy of $\kb \ln \sqrt{2}$ \intext{when two channels are present}, or, \intext{in the case of more --} Fibonacci anyons with $\kb \ln \phi$, where $\phi = \frac{1}{2}(1+\sqrt{5})$ is the golden ratio. A promising experimental realisation for the multi-channel Kondo model are charge Kondo quantum dot devices\cite{Iftikhar2015,Iftikhar2018} \outtext{for} which \outtext{it is predicted that non-Abelian anyons can be detected by entropy measurements.} \intext{hold the potential for detection of non-Abelian anyons through entropy measurements.}\cite{Han2021}

\subsection{Twisted bilayer graphene}
\label{graphene-section}
\intext{Recently electronic entropy measurements have been conducted in twisted bilayer graphene \cite{Rozen2021,Saito2021}. This is a continuous-charge system, so the extension of theory allowing for charge quantisation (Section \ref{quantum-section}) was not required, and instead the Maxwell relation approach outlined in Section \ref{continuous-section} was used.}  

Typically, thermodynamic states with high entropy are more stable at higher temperatures than states with low entropy. \outtext{As an example, a block of solid ice melts when heating it up. Curiously, in}\intext{The} two recent \outtext{experiments} entropy measurements in twisted bilayer graphene\cite{Rozen2021,Saito2021} \outtext{have helped to discover}\intext{revealed} an exotic phenomenon\outtext{ -- the Pomeranchuk effect -- that} \intext{which} seemingly violates this intuitive picture. 
 
\outtext{In both}\intext{Both} studies \intext{measured} the \intext{temperature-dependent} change of the electrochemical potential $\mu$ of one-quarter-filled twisted bilayer graphene,\outtext{ with temperature, has been measured} either directly \cite{Saito2021}, or by integrating the inverse local electronic compressibility\intext{, which was} measured by a nanotube-based scanning single electron transistor\outtext{experiment}\cite{Rozen2021} (see Figure \ref{fig1}a). The entropy of the system could then be determined by using a Maxwell relation (see Equation \ref{maxwell-2}). 

It was found that the entropy per electron of the \intext{high-temperature,} electrically insulating phase \outtext{that is observed at high temperatures} is greater than that of the \intext{low-temperature,} metallic phase\outtext{ found at low temperatures,} by an amount \outtext{that corresponds to that of}\intext{corresponding to} a \intext{single} free \outtext{electron's} spin. \outtext{From this it was concluded }\intext{The suggested explanation for this is} that the high-temperature insulating phase \outtext{possesses}\intext{adopts a} ferromagnetic order, \intext{in which the iso-spins} \outtext{with a low iso-spin}(a combination of valley and spin degree of freedom) \outtext{stiffness, which means that the iso-spins} are globally aligned in one preferred direction, but the constraint on their alignment is weak. \intext{In the metallic phase,} on the other hand, \outtext{in the metallic phase} the constraint on iso-spin alignment is high\outtext{ to ensure} \intext{ensuring} a non-magnetic state \outtext{and thus}\intext{with} an equal number of iso-spins with opposite orientations. Consequently \intext{the stability of the ferromagnetic insulating state is favoured at elevated temperatures due to its slightly higher entropy.} \outtext{This transition -- where electron spins freeze at higher temperature -- is called a Pomeranchuk effect and was first discovered in liquid $^3$He which solidifies when increasing the temperature.}\intext{A similar effect is known as the Pomeranchuk effect in He$_3$, which solidifies when the temperature is increased.}

\subsection{Single electron transistors \intext{and single molecule devices}}
\intext{The recent resurgence in interest towards electronic entropy measurement methods is focused on} single electron transistors \intext{(SETs)} -- devices in which electrical transport is dictated by Coulomb repulsion.\outtext{ and through which electrons flow consequently sequentially (one at a time) --} \intext{SETs} have been realised in \outtext{various}\intext{a wide range of} systems: \outtext{ranging}from electron gases\cite{Meirav90}, semiconductor nanostructures\cite{Reed88}, hetero-nanowires\cite{Thelander03} to single molecule junctions\cite{Liang2002}. \outtext{Thanks to the countable number of charges on such devices, the theoretical framework presented in chapter III can be applied and the entropy of the system's few-electron ground states can be directly extracted from charge transport measurements.}\intext{The theoretical description of quantised-charge entropy measurements in Section \ref{quantum-section} is system-independent and only relies on few charge-states being energetically available.}

\intext{Entropy measurements based on charge sensing were demonstrated as proof of concept in \cite{Hartman2018} (see Figure 1b).} To this end, the entropy \intext{differences} \outtext{of the first, second and third electron ground state in a GaAs quantum dot (Figure 3) has been measured by charge sensing (see Figure 1b) \cite{Hartman2018}.} \intext{were measured for the $N=0\leftrightarrow N=1$ and $N=1 \leftrightarrow N=2$ charge state transitions.} A precise value of $\kb \ln 2$ of a single spin 1/2 was found\outtext{by studying the N=0 to N=1 charge} in the first transition (Figure 3a), \outtext{. Furthermore, the entropy measurements allowed the the authors to observe}\intext{while} a singlet to triplet transition \intext{was observed} in the \outtext{$N=1$ to} $N=2$ \outtext{transition}\intext{charge states} at high magnetic fields (Figure 3b). Such precise entropy measurements \outtext{could be useful} \intext{can be used} for probing more exotic systems like those with non-Abelian statistics (see Section \ref{Majorana-section} "Detection of non-Abelian anyons").

\intext{Molecular electronics}, the concept of using single molecules as building blocks of nanodevices,\outtext{is the core of ‘molecular electronics’ and} gained enormous theoretical and experimental interest over the last decades. \intext{Molecules are fundamentally perfectly reproducible and their }\outtext{Because single molecules enable electronic devices at a}length scales of 1-3~nm\outtext{they} offer\outtext{the a perfect} a platform for \outtext{ultimate} downscaling and\outtext{can} surpass\intext{ing the} limitations of conventional silicon-based technologies\cite{Gehring2019} \intext{by harnessing the achievements of chemical synthesis\cite{Par19, Wid12, Lam16}}.\outtext{Furthermore, many studies have suggested that single molecules hold great potential for enabling applications in energy conversion and solid-state cooling. Thus, their } Thermoelectric \outtext{investigation}\intext{research of molecular devices} has received continuously growing attention in recent years \intext{for potential applications in energy conversion and solid-state cooling} \cite{Reddy2007,Rincon2016,Cui2017,Gehring2017,Gehring2019,Wang2020}. \outtext{Theoretical simulations have shown that the thermoelectric figure of merit $ZT$ can be significantly enhanced by engineering molecule length\cite{Par19}, optimizing the tunnel coupling strength of molecules via chemical anchor groups \cite{Wid12}, or by quantum interference features in their transmission function \cite{Lam16}.} 

\intext{Degeneracy of energy levels in single-molecule devices is of interest, as}\outtext{Furthermore,} it has recently been suggested that molecular systems with highly degenerate ground states \outtext{could lead to}\intext{can lead to} exceptionally high thermoelectric power factors.\cite{Sowa2019} \intext{A conductance-based entropy measurement in}\outtext{Experimental studies on} a Gd-terpyridine complex (Figure 1c), \outtext{where the spin-entropy of the molecular junction was extracted by conductance measurements as}\intext{which followed the method} depicted in Figure 2g, \outtext{could} indeed shows an enhacement of the power factor of the thermoelectric junction by spin entropy, \intext{ originating from energy level degeneracy}.\cite{Gehring2021} 

\intext{Beyond finding strategies to enhance the thermoelectric efficiency,} thermoelectric experiments on single-molecule junctions can furthermore serve as a novel kind of spectroscopy, which allows to \outtext{extract useful information about the spin-ground state of the molecule}\intext{deduce the energy level structure of a molecule from its thermoelectric properties}. \intext{As proof of concept, this was demonstrated }\outtext{To this end, thermocurrent-spectroscopy experiments}on a radical molecule \cite{Pyurbeeva2021b} \intext{(see Fig. \ref{fig3}c,d)} \outtext{revealed that if a second electron is added to the radical molecule a singlet forms at low magnetic field while at higher magnetic fields the molecule is in a triplet ground state} \intext{revealing the presence of a triplet state in the reduced state of the molecule, as well as extracting the energy of the singlet-triplet transition.}\outtext{This singlet-triplet transition could not be observed by using the charge transport data only, highlighting the power of thermocurrent-spectroscopy to quantify the entropy and spin configuration of single-molecular systems.} 

\section{Outlook and Conclusions}
\label{final-section}

\outtext{We have discussed a group of methods that allow to determine entropy differences in a wide range of systems, from bilayer graphene to single molecules, through electric (or thermoelectric) means. Work on electronic entropy measurements started approximately fifteen years ago, but has taken off in the past five.} 
This review has examined the latest chapter in the long history of\outtext{using} entropy measurements as a macroscopic probe for microscopic dynamics. Unlike conventional entropy measurements\outtext{that are} based on\outtext{measuring} heat, these latest measurements probe entropy directly from \outtext{charge and charge transport}\intext{electronic} measurements in \outtext{nanoelectronic devices}\intext{nanodevices}. \outtext{These electronic entropy measurements were}\intext{This idea was} first proposed in the early 2000s and has \outtext{been experimentally demonstrated}\intext{gained attention} in the past few years.

Currently, the \outtext{existing methods are moving on}\intext{field is in transition} from proof-of-concept experiments, \outtext{measuring the entropy of a two-fold degenerate quantum dot}\intext{measuring the entropy of a simple degenerate level} \outtext{(see Figure \ref{fig3}a,b)} \cite{Hartman2018, Kleeorin2019} and conceptual theoretical work \cite{Sela2019} to practical applications, \outtext{such as}\intext{which involved} the demonstration of a Pomeranchuk-type effect in twisted bilayer graphene \cite{Rozen2021, Saito2021} and \outtext{finding}\intext{the measurement of} the singlet-triplet transition energy in a single-molecule device\outtext{(see Figure \ref{fig3}c)}. \intext{In the area of detection of non-Abelian anyons the theoretical background is in place and current work is focused on}\outtext{work on} improving experimental devices for greater \intext{experimental} accuracy \outtext{is also ongoing} \cite{Child2021}.

As\outtext{, and it is simply a matter of time,} electronic entropy measurements complete the transition to yet another tool in the experimentalist's toolbox, they will have two \outtext{possible}\intext{distinct} applications. First, \intext{entropic singatures can be used for detection,} as it is proposed for non-Abelian anyons.\outtext{(see Section \ref{Majorana-section}), measuring entropy can be an indirect signature for the presence of a particular effect, which may not be distinguishable by alternative means.} \intext{Secondly, the value of entropy can allow to}\outtext{Another approach is to used entropy measurements in order to} determine microscopic dynamics \intext{of a system with no a-priori knowledge, as it was demonstrated}\outtext{ which was done to some extent} in twisted bilayer graphene \cite{Rozen2021, Saito2021} \outtext{ (Pomeranchuk effect allows to hypothesise on the magnetic interaction that would lead to it),} and in a single-molecule device \cite{Pyurbeeva2021b}. \outtext{, where entropy measurements led to a full model of the energy level structure in a molecule.} 

\outtext{The latter in particular solves a long-standing problem in molecular electronics, the ultimate goal of which is to utilise achievements in chemical synthesis in order to encode specific electric properties into molecules. The issue is that while standard spectroscopy techniques allow to find energy level structures in free molecules in a solution, it is largely unknown how coupling to metallic leads and incorporation into a solid-state device affects this structure. Thermoelectric spectroscopy allows to find the energy level structure directly in situ, especially if the model is known from standard measurements in a solution, and only the charges of parameters need to be determined. }

In this review we have attempted to develop a taxonomy for \outtext{the different}\intext{existing} electronic measurement techniques, and \intext{to} reveal their underlying thermodynamic principles, \intext{as well as their limitations.}

\outtext{A strength of transport-based entropy measurement methods is that they can be performed on devices that are otherwise widely studied, usually for thermoelectric research. In some cases \cite{Kim2014, Dorsch2021}, they can even be used as a post-factum analysis of existing experimental data. }

\intext{The} consideration of limitations \outtext{of entropy measurement methods (Section \ref{limitations-section})}\intext{itself} opens up new questions and areas of research. Examples of these include studying mechanical degrees of freedom in \outtext{small}systems with strong electron-vibrational coupling and\outtext{electronic properties of} devices in the rarely-considered intermediate case between the Landauer regime and weakly-coupled sequential tunnelling, in which lifetime in the device is small and has a significant contribution to entropy \cite{Child2021a}.  

Finally, \outtext{being able}\intext{the ability} to measure thermodynamic parameters, primarily entropy, in nanodevices \intext{opens them up as an experimental platform for studying}\outtext{in which effects of} stochastic, and especially quantum thermodynamics. \outtext{ play a large role would allow to use nanodevices for experimental study of both.}\intext{This is in contrast with the usual bottom-up experimental approach to the field \cite{Micadei2017}, but has the}\outtext{The usual approach taken experimentally is more bottom-up, including precise quantum manipulation of relatively few parameters and has been able to demonstrate the most unusual predictions of quantum thermodynamics \cite{Micadei2017}. While such precision can not be achieved in electric nanodevices, they have the} benefit of years of technological advancement \intext{of nanodevices} for practical applications, \outtext{and}\intext{as well as} ease of experimentation. \intext{This approach}\outtext{The last application} would require \outtext{some}\intext{additional} theoretical work, as the term “open quantum systems” usually implies systems open to energy, but not particle exchange. However the experimental accessibility of electrical nanodevices which include both\outtext{, unlike, for example, molecular rotors, can justify} \intext{justifies} this effort. 

\intext{In summary,} to date, \outtext{most effort has been}\intext{the field has focused} on developing\outtext{these} novel measurement techniques\outtext{, with}\intext{and conceptual} demonstration of entropy measurements in a \outtext{host of materials systems and devices}\intext{range of systems}. \intext{Moving forwards,} the next stage will be to apply these techniques in a systematic way to uncover microscopic dynamics of quantum correlated systems, \outtext{and}potentially discover new and exotic quasi-particles in condensed matter systems, \intext{as well as expand the theoretical background for more fundamental applications}.

 \outtext{Much like heat-based measurements of entropy enabled the development of engines that powered a technological revolution in the 19$^{\text{th}}$ century, electronic measurements of entropy in meso- and nanoscale systems will power the quantum technology revolution of the 21$^{\text{st}}$ century.}

\section*{Acknowledgements}
 P.G. acknowledges financial support from the F.R.S.-FNRS of Belgium (FNRS-CQ-1.C044.21-SMARD, FNRS-CDR-J.0068.21-SMARD, FNRS-MIS-F.4523.22-TopoBrain), from the Federation Wallonie-Bruxelles through the ARC Grant No. 21/26-116 and from the EU (ERC-StG-10104144-MOUNTAIN). This project (40007563-CONNECT) has received funding from the FWO and F.R.S.-FNRS under the Excellence of Science (EOS) programme.
 
 J.A.M. was supported through the UKRI Future Leaders Fellowship, Grant No. MR/S032541/1, with in-kind support from the Royal Academy of Engineering.


%
%

%



\begin{thebibliography}{86}%
\makeatletter
\providecommand \@ifxundefined [1]{%
 \@ifx{#1\undefined}
}%
\providecommand \@ifnum [1]{%
 \ifnum #1\expandafter \@firstoftwo
 \else \expandafter \@secondoftwo
 \fi
}%
\providecommand \@ifx [1]{%
 \ifx #1\expandafter \@firstoftwo
 \else \expandafter \@secondoftwo
 \fi
}%
\providecommand \natexlab [1]{#1}%
\providecommand \enquote  [1]{``#1''}%
\providecommand \bibnamefont  [1]{#1}%
\providecommand \bibfnamefont [1]{#1}%
\providecommand \citenamefont [1]{#1}%
\providecommand \href@noop [0]{\@secondoftwo}%
\providecommand \href [0]{\begingroup \@sanitize@url \@href}%
\providecommand \@href[1]{\@@startlink{#1}\@@href}%
\providecommand \@@href[1]{\endgroup#1\@@endlink}%
\providecommand \@sanitize@url [0]{\catcode `\\12\catcode `\$12\catcode
  `\&12\catcode `\#12\catcode `\^12\catcode `\_12\catcode `\%12\relax}%
\providecommand \@@startlink[1]{}%
\providecommand \@@endlink[0]{}%
\providecommand \url  [0]{\begingroup\@sanitize@url \@url }%
\providecommand \@url [1]{\endgroup\@href {#1}{\urlprefix }}%
\providecommand \urlprefix  [0]{URL }%
\providecommand \Eprint [0]{\href }%
\providecommand \doibase [0]{http://dx.doi.org/}%
\providecommand \selectlanguage [0]{\@gobble}%
\providecommand \bibinfo  [0]{\@secondoftwo}%
\providecommand \bibfield  [0]{\@secondoftwo}%
\providecommand \translation [1]{[#1]}%
\providecommand \BibitemOpen [0]{}%
\providecommand \bibitemStop [0]{}%
\providecommand \bibitemNoStop [0]{.\EOS\space}%
\providecommand \EOS [0]{\spacefactor3000\relax}%
\providecommand \BibitemShut  [1]{\csname bibitem#1\endcsname}%
\let\auto@bib@innerbib\@empty
\bibitem [{\citenamefont {Maxwell}(1860)}]{Maxwell1860}%
  \BibitemOpen
  \bibfield  {author} {\bibinfo {author} {\bibfnamefont {J.~C.}\ \bibnamefont
  {Maxwell}},\ }\bibfield  {title} {\enquote {\bibinfo {title} {{II.
  Illustrations of the dynamical theory of gases}},}\ }\href {\doibase
  10.1080/14786446008642902} {\bibfield  {journal} {\bibinfo  {journal}
  {London, Edinburgh, Dublin Philos. Mag. J. Sci.}\ }\textbf {\bibinfo {volume}
  {20}},\ \bibinfo {pages} {21--37} (\bibinfo {year} {1860})}\BibitemShut
  {NoStop}%
\bibitem [{\citenamefont {Bragg}\ and\ \citenamefont
  {Williams}(1934)}]{Bragg1934}%
  \BibitemOpen
  \bibfield  {author} {\bibinfo {author} {\bibfnamefont {W.~L.}\ \bibnamefont
  {Bragg}}\ and\ \bibinfo {author} {\bibfnamefont {E.~J.}\ \bibnamefont
  {Williams}},\ }\bibfield  {title} {\enquote {\bibinfo {title} {{The effect of
  thermal agitation on atomic arrangement in alloys}},}\ }\href {\doibase
  10.1098/rspa.1934.0132} {\bibfield  {journal} {\bibinfo  {journal} {Proc. R.
  Soc. London. Ser. A, Contain. Pap. a Math. Phys. Character}\ }\textbf
  {\bibinfo {volume} {145}},\ \bibinfo {pages} {699--730} (\bibinfo {year}
  {1934})}\BibitemShut {NoStop}%
\bibitem [{\citenamefont {Bragg}\ and\ \citenamefont
  {Williams}(1935)}]{Bragg1935}%
  \BibitemOpen
  \bibfield  {author} {\bibinfo {author} {\bibfnamefont {W.~L.}\ \bibnamefont
  {Bragg}}\ and\ \bibinfo {author} {\bibfnamefont {E.~J.}\ \bibnamefont
  {Williams}},\ }\bibfield  {title} {\enquote {\bibinfo {title} {{The effect of
  thermal agitation on atomic arrangement in alloys-III}},}\ }\href {\doibase
  10.1098/rspa.1935.0188} {\bibfield  {journal} {\bibinfo  {journal} {Proc. R.
  Soc. London. Ser. A - Math. Phys. Sci.}\ }\textbf {\bibinfo {volume} {152}},\
  \bibinfo {pages} {231--252} (\bibinfo {year} {1935})}\BibitemShut {NoStop}%
\bibitem [{\citenamefont {Nix}\ and\ \citenamefont {Shockley}(1938)}]{Nix1938}%
  \BibitemOpen
  \bibfield  {author} {\bibinfo {author} {\bibfnamefont {F.~C.}\ \bibnamefont
  {Nix}}\ and\ \bibinfo {author} {\bibfnamefont {W.}~\bibnamefont {Shockley}},\
  }\bibfield  {title} {\enquote {\bibinfo {title} {{Order-Disorder
  Transformations in Alloys}},}\ }\href {\doibase 10.1103/RevModPhys.10.1}
  {\bibfield  {journal} {\bibinfo  {journal} {Rev. Mod. Phys.}\ }\textbf
  {\bibinfo {volume} {10}},\ \bibinfo {pages} {1--71} (\bibinfo {year}
  {1938})}\BibitemShut {NoStop}%
\bibitem [{\citenamefont {Guthrie}\ and\ \citenamefont
  {McCullough}(1961)}]{Guthrie1961}%
  \BibitemOpen
  \bibfield  {author} {\bibinfo {author} {\bibfnamefont {G.~B.}\ \bibnamefont
  {Guthrie}}\ and\ \bibinfo {author} {\bibfnamefont {J.~P.}\ \bibnamefont
  {McCullough}},\ }\bibfield  {title} {\enquote {\bibinfo {title} {{Some
  observations on phase transformations in molecular crystals}},}\ }\href
  {\doibase 10.1016/0022-3697(61)90083-X} {\bibfield  {journal} {\bibinfo
  {journal} {J. Phys. Chem. Solids}\ }\textbf {\bibinfo {volume} {18}},\
  \bibinfo {pages} {53--61} (\bibinfo {year} {1961})}\BibitemShut {NoStop}%
\bibitem [{\citenamefont {Ramirez}\ \emph {et~al.}(1999)\citenamefont
  {Ramirez}, \citenamefont {Hayashi}, \citenamefont {Cava}, \citenamefont
  {Siddharthan},\ and\ \citenamefont {Shastry}}]{Ramirez1999}%
  \BibitemOpen
  \bibfield  {author} {\bibinfo {author} {\bibfnamefont {A.~P.}\ \bibnamefont
  {Ramirez}}, \bibinfo {author} {\bibfnamefont {A.}~\bibnamefont {Hayashi}},
  \bibinfo {author} {\bibfnamefont {R.~J.}\ \bibnamefont {Cava}}, \bibinfo
  {author} {\bibfnamefont {R.}~\bibnamefont {Siddharthan}}, \ and\ \bibinfo
  {author} {\bibfnamefont {B.~S.}\ \bibnamefont {Shastry}},\ }\bibfield
  {title} {\enquote {\bibinfo {title} {{Zero-point entropy in 'spin ice'}},}\
  }\href {\doibase 10.1038/20619} {\bibfield  {journal} {\bibinfo  {journal}
  {Nature}\ }\textbf {\bibinfo {volume} {399}},\ \bibinfo {pages} {333--335}
  (\bibinfo {year} {1999})}\BibitemShut {NoStop}%
\bibitem [{\citenamefont {Gornik}\ \emph {et~al.}(1985)\citenamefont {Gornik},
  \citenamefont {Lassnig}, \citenamefont {Strasser}, \citenamefont
  {St{\"{o}}rmer}, \citenamefont {Gossard},\ and\ \citenamefont
  {Wiegmann}}]{Gornik1985}%
  \BibitemOpen
  \bibfield  {author} {\bibinfo {author} {\bibfnamefont {E.}~\bibnamefont
  {Gornik}}, \bibinfo {author} {\bibfnamefont {R.}~\bibnamefont {Lassnig}},
  \bibinfo {author} {\bibfnamefont {G.}~\bibnamefont {Strasser}}, \bibinfo
  {author} {\bibfnamefont {H.~L.}\ \bibnamefont {St{\"{o}}rmer}}, \bibinfo
  {author} {\bibfnamefont {A.~C.}\ \bibnamefont {Gossard}}, \ and\ \bibinfo
  {author} {\bibfnamefont {W.}~\bibnamefont {Wiegmann}},\ }\bibfield  {title}
  {\enquote {\bibinfo {title} {{Specific heat of two-dimensional electrons in
  GaAs-GaAlAs multilayers}},}\ }\href {\doibase 10.1103/PhysRevLett.54.1820}
  {\bibfield  {journal} {\bibinfo  {journal} {Phys. Rev. Lett.}\ }\textbf
  {\bibinfo {volume} {54}},\ \bibinfo {pages} {1820--1823} (\bibinfo {year}
  {1985})}\BibitemShut {NoStop}%
\bibitem [{\citenamefont {Wang}\ \emph {et~al.}(1988)\citenamefont {Wang},
  \citenamefont {Campbell}, \citenamefont {Tsui},\ and\ \citenamefont
  {Cho}}]{Wang1988}%
  \BibitemOpen
  \bibfield  {author} {\bibinfo {author} {\bibfnamefont {J.~K.}\ \bibnamefont
  {Wang}}, \bibinfo {author} {\bibfnamefont {J.~H.}\ \bibnamefont {Campbell}},
  \bibinfo {author} {\bibfnamefont {D.~C.}\ \bibnamefont {Tsui}}, \ and\
  \bibinfo {author} {\bibfnamefont {A.~Y.}\ \bibnamefont {Cho}},\ }\bibfield
  {title} {\enquote {\bibinfo {title} {{Heat capacity of the two-dimensional
  electron gas in GaAs/AlxGa1-xAs multiple-quantum-well structures}},}\ }\href
  {\doibase 10.1103/PhysRevB.38.6174} {\bibfield  {journal} {\bibinfo
  {journal} {Phys. Rev. B}\ }\textbf {\bibinfo {volume} {38}},\ \bibinfo
  {pages} {6174--6184} (\bibinfo {year} {1988})}\BibitemShut {NoStop}%
\bibitem [{\citenamefont {Bayot}\ \emph {et~al.}(1996)\citenamefont {Bayot},
  \citenamefont {Grivei}, \citenamefont {Melinte}, \citenamefont {Santos},\
  and\ \citenamefont {Shayegan}}]{Bayot1996}%
  \BibitemOpen
  \bibfield  {author} {\bibinfo {author} {\bibfnamefont {V.}~\bibnamefont
  {Bayot}}, \bibinfo {author} {\bibfnamefont {E.}~\bibnamefont {Grivei}},
  \bibinfo {author} {\bibfnamefont {S.}~\bibnamefont {Melinte}}, \bibinfo
  {author} {\bibfnamefont {M.~B.}\ \bibnamefont {Santos}}, \ and\ \bibinfo
  {author} {\bibfnamefont {M.}~\bibnamefont {Shayegan}},\ }\bibfield  {title}
  {\enquote {\bibinfo {title} {{Giant low temperature heat capacity of GaAs
  quantum wells near landau level filling $\nu$ = 1}},}\ }\href {\doibase
  10.1103/PhysRevLett.76.4584} {\bibfield  {journal} {\bibinfo  {journal}
  {Phys. Rev. Lett.}\ }\textbf {\bibinfo {volume} {76}},\ \bibinfo {pages}
  {4584--4587} (\bibinfo {year} {1996})}\BibitemShut {NoStop}%
\bibitem [{\citenamefont {Schulze-Wischeler}\ \emph {et~al.}(2007)\citenamefont
  {Schulze-Wischeler}, \citenamefont {Zeitler}, \citenamefont {v.~Zobeltitz},
  \citenamefont {Hohls}, \citenamefont {Reuter}, \citenamefont {Wieck},
  \citenamefont {Frahm},\ and\ \citenamefont {Haug}}]{Schulze-Wischeler2007}%
  \BibitemOpen
  \bibfield  {author} {\bibinfo {author} {\bibfnamefont {F.}~\bibnamefont
  {Schulze-Wischeler}}, \bibinfo {author} {\bibfnamefont {U.}~\bibnamefont
  {Zeitler}}, \bibinfo {author} {\bibfnamefont {C.}~\bibnamefont
  {v.~Zobeltitz}}, \bibinfo {author} {\bibfnamefont {F.}~\bibnamefont {Hohls}},
  \bibinfo {author} {\bibfnamefont {D.}~\bibnamefont {Reuter}}, \bibinfo
  {author} {\bibfnamefont {A.~D.}\ \bibnamefont {Wieck}}, \bibinfo {author}
  {\bibfnamefont {H.}~\bibnamefont {Frahm}}, \ and\ \bibinfo {author}
  {\bibfnamefont {R.~J.}\ \bibnamefont {Haug}},\ }\bibfield  {title} {\enquote
  {\bibinfo {title} {{Measurement of the specific heat of a fractional quantum
  Hall system}},}\ }\href {\doibase 10.1103/PhysRevB.76.153311} {\bibfield
  {journal} {\bibinfo  {journal} {Phys. Rev. B}\ }\textbf {\bibinfo {volume}
  {76}},\ \bibinfo {pages} {153311} (\bibinfo {year} {2007})}\BibitemShut
  {NoStop}%
\bibitem [{\citenamefont {Schmidt}\ \emph {et~al.}(2017)\citenamefont
  {Schmidt}, \citenamefont {Bennaceur}, \citenamefont {Gaucher}, \citenamefont
  {Gervais}, \citenamefont {Pfeiffer},\ and\ \citenamefont
  {West}}]{Schmidt2017}%
  \BibitemOpen
  \bibfield  {author} {\bibinfo {author} {\bibfnamefont {B.~A.}\ \bibnamefont
  {Schmidt}}, \bibinfo {author} {\bibfnamefont {K.}~\bibnamefont {Bennaceur}},
  \bibinfo {author} {\bibfnamefont {S.}~\bibnamefont {Gaucher}}, \bibinfo
  {author} {\bibfnamefont {G.}~\bibnamefont {Gervais}}, \bibinfo {author}
  {\bibfnamefont {L.~N.}\ \bibnamefont {Pfeiffer}}, \ and\ \bibinfo {author}
  {\bibfnamefont {K.~W.}\ \bibnamefont {West}},\ }\bibfield  {title} {\enquote
  {\bibinfo {title} {{Specific heat and entropy of fractional quantum Hall
  states in the second Landau level}},}\ }\href {\doibase
  10.1103/PhysRevB.95.201306} {\bibfield  {journal} {\bibinfo  {journal} {Phys.
  Rev. B}\ }\textbf {\bibinfo {volume} {95}},\ \bibinfo {pages} {1--5}
  (\bibinfo {year} {2017})},\ \Eprint {http://arxiv.org/abs/1605.02344}
  {arXiv:1605.02344} \BibitemShut {NoStop}%
\bibitem [{\citenamefont {Potts}()}]{Potts}%
  \BibitemOpen
  \bibfield  {author} {\bibinfo {author} {\bibfnamefont {P.~P.}\ \bibnamefont
  {Potts}},\ }\href {https://arxiv.org/pdf/1906.07439.pdf} {\enquote {\bibinfo
  {title} {{Introduction to Quantum Thermodynamics}},}\ }\bibinfo {type} {Tech.
  Rep.},\ \Eprint {http://arxiv.org/abs/1906.07439v1} {arXiv:1906.07439v1}
  \BibitemShut {NoStop}%
\bibitem [{\citenamefont {Kosloff}(2013)}]{Kosloff2013a}%
  \BibitemOpen
  \bibfield  {author} {\bibinfo {author} {\bibfnamefont {R.}~\bibnamefont
  {Kosloff}},\ }\bibfield  {title} {\enquote {\bibinfo {title} {{Quantum
  Thermodynamics: A Dynamical Viewpoint}},}\ }\href {\doibase
  10.3390/e15062100} {\bibfield  {journal} {\bibinfo  {journal} {Entropy}\
  }\textbf {\bibinfo {volume} {15}},\ \bibinfo {pages} {2100--2128} (\bibinfo
  {year} {2013})},\ \Eprint {http://arxiv.org/abs/arXiv:1305.2268v1}
  {arXiv:arXiv:1305.2268v1} \BibitemShut {NoStop}%
\bibitem [{\citenamefont {Rozen}\ \emph {et~al.}(2021)\citenamefont {Rozen},
  \citenamefont {Park}, \citenamefont {Zondiner}, \citenamefont {Cao},
  \citenamefont {Rodan-Legrain}, \citenamefont {Taniguchi}, \citenamefont
  {Watanabe}, \citenamefont {Oreg}, \citenamefont {Stern}, \citenamefont
  {Berg}, \citenamefont {Jarillo-Herrero},\ and\ \citenamefont
  {Ilani}}]{Rozen2021}%
  \BibitemOpen
  \bibfield  {author} {\bibinfo {author} {\bibfnamefont {A.}~\bibnamefont
  {Rozen}}, \bibinfo {author} {\bibfnamefont {J.~M.}\ \bibnamefont {Park}},
  \bibinfo {author} {\bibfnamefont {U.}~\bibnamefont {Zondiner}}, \bibinfo
  {author} {\bibfnamefont {Y.}~\bibnamefont {Cao}}, \bibinfo {author}
  {\bibfnamefont {D.}~\bibnamefont {Rodan-Legrain}}, \bibinfo {author}
  {\bibfnamefont {T.}~\bibnamefont {Taniguchi}}, \bibinfo {author}
  {\bibfnamefont {K.}~\bibnamefont {Watanabe}}, \bibinfo {author}
  {\bibfnamefont {Y.}~\bibnamefont {Oreg}}, \bibinfo {author} {\bibfnamefont
  {A.}~\bibnamefont {Stern}}, \bibinfo {author} {\bibfnamefont
  {E.}~\bibnamefont {Berg}}, \bibinfo {author} {\bibfnamefont {P.}~\bibnamefont
  {Jarillo-Herrero}}, \ and\ \bibinfo {author} {\bibfnamefont {S.}~\bibnamefont
  {Ilani}},\ }\bibfield  {title} {\enquote {\bibinfo {title} {Entropic evidence
  for a pomeranchuk effect in magic-angle graphene},}\ }\href {\doibase
  10.1038/s41586-021-03319-3} {\bibfield  {journal} {\bibinfo  {journal}
  {Nature}\ }\textbf {\bibinfo {volume} {592}},\ \bibinfo {pages} {214--219}
  (\bibinfo {year} {2021})}\BibitemShut {NoStop}%
\bibitem [{\citenamefont {Saito}\ \emph {et~al.}(2021)\citenamefont {Saito},
  \citenamefont {Yang}, \citenamefont {Ge}, \citenamefont {Liu}, \citenamefont
  {Taniguchi}, \citenamefont {Watanabe}, \citenamefont {Li}, \citenamefont
  {Berg},\ and\ \citenamefont {Young}}]{Saito2021}%
  \BibitemOpen
  \bibfield  {author} {\bibinfo {author} {\bibfnamefont {Y.}~\bibnamefont
  {Saito}}, \bibinfo {author} {\bibfnamefont {F.}~\bibnamefont {Yang}},
  \bibinfo {author} {\bibfnamefont {J.}~\bibnamefont {Ge}}, \bibinfo {author}
  {\bibfnamefont {X.}~\bibnamefont {Liu}}, \bibinfo {author} {\bibfnamefont
  {T.}~\bibnamefont {Taniguchi}}, \bibinfo {author} {\bibfnamefont
  {K.}~\bibnamefont {Watanabe}}, \bibinfo {author} {\bibfnamefont {J.~I.~A.}\
  \bibnamefont {Li}}, \bibinfo {author} {\bibfnamefont {E.}~\bibnamefont
  {Berg}}, \ and\ \bibinfo {author} {\bibfnamefont {A.~F.}\ \bibnamefont
  {Young}},\ }\bibfield  {title} {\enquote {\bibinfo {title} {Isospin
  pomeranchuk effect in twisted bilayer graphene},}\ }\href {\doibase
  10.1038/s41586-021-03409-2} {\bibfield  {journal} {\bibinfo  {journal}
  {Nature}\ }\textbf {\bibinfo {volume} {592}},\ \bibinfo {pages} {220--224}
  (\bibinfo {year} {2021})}\BibitemShut {NoStop}%
\bibitem [{\citenamefont {Han}\ \emph {et~al.}(2021)\citenamefont {Han},
  \citenamefont {Mitchell}, \citenamefont {Iftikhar}, \citenamefont {Kleeorin},
  \citenamefont {Anthore}, \citenamefont {Pierre}, \citenamefont {Meir},\ and\
  \citenamefont {Sela}}]{Han2021}%
  \BibitemOpen
  \bibfield  {author} {\bibinfo {author} {\bibfnamefont {C.}~\bibnamefont
  {Han}}, \bibinfo {author} {\bibfnamefont {A.~K.}\ \bibnamefont {Mitchell}},
  \bibinfo {author} {\bibfnamefont {Z.}~\bibnamefont {Iftikhar}}, \bibinfo
  {author} {\bibfnamefont {Y.}~\bibnamefont {Kleeorin}}, \bibinfo {author}
  {\bibfnamefont {A.}~\bibnamefont {Anthore}}, \bibinfo {author} {\bibfnamefont
  {F.}~\bibnamefont {Pierre}}, \bibinfo {author} {\bibfnamefont
  {Y.}~\bibnamefont {Meir}}, \ and\ \bibinfo {author} {\bibfnamefont
  {E.}~\bibnamefont {Sela}},\ }\bibfield  {title} {\enquote {\bibinfo {title}
  {{Fractional entropy of multichannel Kondo systems from conductance-charge
  relations}},}\ }\href {http://arxiv.org/abs/2108.12878} {\ \textbf {\bibinfo
  {volume} {2}} (\bibinfo {year} {2021})},\ \Eprint
  {http://arxiv.org/abs/2108.12878} {arXiv:2108.12878} \BibitemShut {NoStop}%
\bibitem [{\citenamefont {Hartman}\ \emph {et~al.}(2018)\citenamefont
  {Hartman}, \citenamefont {Olsen}, \citenamefont {L{\"u}scher}, \citenamefont
  {Samani}, \citenamefont {Fallahi}, \citenamefont {Gardner}, \citenamefont
  {Manfra},\ and\ \citenamefont {Folk}}]{Hartman2018}%
  \BibitemOpen
  \bibfield  {author} {\bibinfo {author} {\bibfnamefont {N.}~\bibnamefont
  {Hartman}}, \bibinfo {author} {\bibfnamefont {C.}~\bibnamefont {Olsen}},
  \bibinfo {author} {\bibfnamefont {S.}~\bibnamefont {L{\"u}scher}}, \bibinfo
  {author} {\bibfnamefont {M.}~\bibnamefont {Samani}}, \bibinfo {author}
  {\bibfnamefont {S.}~\bibnamefont {Fallahi}}, \bibinfo {author} {\bibfnamefont
  {G.~C.}\ \bibnamefont {Gardner}}, \bibinfo {author} {\bibfnamefont
  {M.}~\bibnamefont {Manfra}}, \ and\ \bibinfo {author} {\bibfnamefont
  {J.}~\bibnamefont {Folk}},\ }\bibfield  {title} {\enquote {\bibinfo {title}
  {Direct entropy measurement in a mesoscopic quantum system},}\ }\href
  {\doibase 10.1038/s41567-018-0250-5} {\bibfield  {journal} {\bibinfo
  {journal} {Nature Physics}\ }\textbf {\bibinfo {volume} {14}},\ \bibinfo
  {pages} {1083--1086} (\bibinfo {year} {2018})}\BibitemShut {NoStop}%
\bibitem [{\citenamefont {Kleeorin}\ \emph {et~al.}(2019)\citenamefont
  {Kleeorin}, \citenamefont {Thierschmann}, \citenamefont {Buhmann},
  \citenamefont {Georges}, \citenamefont {Molenkamp},\ and\ \citenamefont
  {Meir}}]{Kleeorin2019}%
  \BibitemOpen
  \bibfield  {author} {\bibinfo {author} {\bibfnamefont {Y.}~\bibnamefont
  {Kleeorin}}, \bibinfo {author} {\bibfnamefont {H.}~\bibnamefont
  {Thierschmann}}, \bibinfo {author} {\bibfnamefont {H.}~\bibnamefont
  {Buhmann}}, \bibinfo {author} {\bibfnamefont {A.}~\bibnamefont {Georges}},
  \bibinfo {author} {\bibfnamefont {L.~W.}\ \bibnamefont {Molenkamp}}, \ and\
  \bibinfo {author} {\bibfnamefont {Y.}~\bibnamefont {Meir}},\ }\bibfield
  {title} {\enquote {\bibinfo {title} {{How to measure the entropy of a
  mesoscopic system via thermoelectric transport}},}\ }\href {\doibase
  10.1038/s41467-019-13630-3} {\bibfield  {journal} {\bibinfo  {journal} {Nat.
  Commun.}\ }\textbf {\bibinfo {volume} {10}},\ \bibinfo {pages} {5801}
  (\bibinfo {year} {2019})},\ \Eprint {http://arxiv.org/abs/1904.08948}
  {arXiv:1904.08948} \BibitemShut {NoStop}%
\bibitem [{\citenamefont {Gehring}\ \emph {et~al.}(2021)\citenamefont
  {Gehring}, \citenamefont {Sowa}, \citenamefont {Hsu}, \citenamefont
  {de~Bruijckere}, \citenamefont {van~der Star}, \citenamefont {{Le Roy}},
  \citenamefont {Bogani}, \citenamefont {Gauger},\ and\ \citenamefont {van~der
  Zant}}]{Gehring2021}%
  \BibitemOpen
  \bibfield  {author} {\bibinfo {author} {\bibfnamefont {P.}~\bibnamefont
  {Gehring}}, \bibinfo {author} {\bibfnamefont {J.~K.}\ \bibnamefont {Sowa}},
  \bibinfo {author} {\bibfnamefont {C.}~\bibnamefont {Hsu}}, \bibinfo {author}
  {\bibfnamefont {J.}~\bibnamefont {de~Bruijckere}}, \bibinfo {author}
  {\bibfnamefont {M.}~\bibnamefont {van~der Star}}, \bibinfo {author}
  {\bibfnamefont {J.~J.}\ \bibnamefont {{Le Roy}}}, \bibinfo {author}
  {\bibfnamefont {L.}~\bibnamefont {Bogani}}, \bibinfo {author} {\bibfnamefont
  {E.~M.}\ \bibnamefont {Gauger}}, \ and\ \bibinfo {author} {\bibfnamefont
  {H.~S.~J.}\ \bibnamefont {van~der Zant}},\ }\bibfield  {title} {\enquote
  {\bibinfo {title} {{Complete mapping of the thermoelectric properties of a
  single molecule}},}\ }\href {\doibase 10.1038/s41565-021-00859-7} {\bibfield
  {journal} {\bibinfo  {journal} {Nat. Nanotechnol.}\ }\textbf {\bibinfo
  {volume} {16}},\ \bibinfo {pages} {426--430} (\bibinfo {year}
  {2021})}\BibitemShut {NoStop}%
\bibitem [{\citenamefont {Pyurbeeva}\ and\ \citenamefont
  {Mol}(2021)}]{Pyurbeeva2021}%
  \BibitemOpen
  \bibfield  {author} {\bibinfo {author} {\bibfnamefont {E.}~\bibnamefont
  {Pyurbeeva}}\ and\ \bibinfo {author} {\bibfnamefont {J.~A.}\ \bibnamefont
  {Mol}},\ }\bibfield  {title} {\enquote {\bibinfo {title} {{A Thermodynamic
  Approach to Measuring Entropy in a Few-Electron Nanodevice}},}\ }\href
  {\doibase 10.3390/e23060640} {\bibfield  {journal} {\bibinfo  {journal}
  {Entropy}\ }\textbf {\bibinfo {volume} {23}},\ \bibinfo {pages} {640}
  (\bibinfo {year} {2021})},\ \Eprint {http://arxiv.org/abs/2008.05747}
  {arXiv:2008.05747} \BibitemShut {NoStop}%
\bibitem [{\citenamefont {Josefsson}\ \emph {et~al.}(2018)\citenamefont
  {Josefsson}, \citenamefont {Svilans}, \citenamefont {Burke}, \citenamefont
  {Hoffmann}, \citenamefont {Fahlvik}, \citenamefont {Thelander}, \citenamefont
  {Leijnse},\ and\ \citenamefont {Linke}}]{Josefssonb}%
  \BibitemOpen
  \bibfield  {author} {\bibinfo {author} {\bibfnamefont {M.}~\bibnamefont
  {Josefsson}}, \bibinfo {author} {\bibfnamefont {A.}~\bibnamefont {Svilans}},
  \bibinfo {author} {\bibfnamefont {A.~M.}\ \bibnamefont {Burke}}, \bibinfo
  {author} {\bibfnamefont {E.~A.}\ \bibnamefont {Hoffmann}}, \bibinfo {author}
  {\bibfnamefont {S.}~\bibnamefont {Fahlvik}}, \bibinfo {author} {\bibfnamefont
  {C.}~\bibnamefont {Thelander}}, \bibinfo {author} {\bibfnamefont
  {M.}~\bibnamefont {Leijnse}}, \ and\ \bibinfo {author} {\bibfnamefont
  {H.}~\bibnamefont {Linke}},\ }\bibfield  {title} {\enquote {\bibinfo {title}
  {{A quantum-dot heat engine operating close to the thermodynamic efficiency
  limits}},}\ }\href {\doibase 10.1038/s41565-018-0200-5} {\bibfield  {journal}
  {\bibinfo  {journal} {Nat. Nanotechnol.}\ }\textbf {\bibinfo {volume} {13}},\
  \bibinfo {pages} {920--924} (\bibinfo {year} {2018})}\BibitemShut {NoStop}%
\bibitem [{\citenamefont {Koski}\ \emph {et~al.}(2015)\citenamefont {Koski},
  \citenamefont {Kutvonen}, \citenamefont {Khaymovich}, \citenamefont
  {Ala-Nissila},\ and\ \citenamefont {Pekola}}]{Koski2015}%
  \BibitemOpen
  \bibfield  {author} {\bibinfo {author} {\bibfnamefont {J.~V.}\ \bibnamefont
  {Koski}}, \bibinfo {author} {\bibfnamefont {A.}~\bibnamefont {Kutvonen}},
  \bibinfo {author} {\bibfnamefont {I.~M.}\ \bibnamefont {Khaymovich}},
  \bibinfo {author} {\bibfnamefont {T.}~\bibnamefont {Ala-Nissila}}, \ and\
  \bibinfo {author} {\bibfnamefont {J.~P.}\ \bibnamefont {Pekola}},\ }\bibfield
   {title} {\enquote {\bibinfo {title} {{On-Chip Maxwell's Demon as an
  Information-Powered Refrigerator}},}\ }\href {\doibase
  10.1103/PhysRevLett.115.260602} {\  (\bibinfo {year} {2015}),\
  10.1103/PhysRevLett.115.260602}\BibitemShut {NoStop}%
\bibitem [{\citenamefont {S{\'{a}}nchez}, \citenamefont {Samuelsson},\ and\
  \citenamefont {Potts}(2019)}]{Sanchez2019}%
  \BibitemOpen
  \bibfield  {author} {\bibinfo {author} {\bibfnamefont {R.}~\bibnamefont
  {S{\'{a}}nchez}}, \bibinfo {author} {\bibfnamefont {P.}~\bibnamefont
  {Samuelsson}}, \ and\ \bibinfo {author} {\bibfnamefont {P.~P.}\ \bibnamefont
  {Potts}},\ }\bibfield  {title} {\enquote {\bibinfo {title} {{Autonomous
  conversion of information to work in quantum dots}},}\ }\href
  {http://arxiv.org/abs/1907.02866} {\ ,\ \bibinfo {pages} {1--18} (\bibinfo
  {year} {2019})},\ \Eprint {http://arxiv.org/abs/1907.02866}
  {arXiv:1907.02866} \BibitemShut {NoStop}%
\bibitem [{\citenamefont {S{\'{a}}nchez}, \citenamefont {Splettstoesser},\ and\
  \citenamefont {Whitney}(2019)}]{Sanchez2019a}%
  \BibitemOpen
  \bibfield  {author} {\bibinfo {author} {\bibfnamefont {R.}~\bibnamefont
  {S{\'{a}}nchez}}, \bibinfo {author} {\bibfnamefont {J.}~\bibnamefont
  {Splettstoesser}}, \ and\ \bibinfo {author} {\bibfnamefont {R.~S.}\
  \bibnamefont {Whitney}},\ }\bibfield  {title} {\enquote {\bibinfo {title}
  {{Nonequilibrium System as a Demon}},}\ }\href {\doibase
  10.1103/PhysRevLett.123.216801} {\bibfield  {journal} {\bibinfo  {journal}
  {Phys. Rev. Lett.}\ }\textbf {\bibinfo {volume} {123}},\ \bibinfo {pages}
  {216801} (\bibinfo {year} {2019})},\ \Eprint
  {http://arxiv.org/abs/1811.02453} {arXiv:1811.02453} \BibitemShut {NoStop}%
\bibitem [{\citenamefont {Kim}\ \emph {et~al.}(2011)\citenamefont {Kim},
  \citenamefont {Sagawa}, \citenamefont {{De Liberato}},\ and\ \citenamefont
  {Ueda}}]{Kim2011}%
  \BibitemOpen
  \bibfield  {author} {\bibinfo {author} {\bibfnamefont {S.~W.}\ \bibnamefont
  {Kim}}, \bibinfo {author} {\bibfnamefont {T.}~\bibnamefont {Sagawa}},
  \bibinfo {author} {\bibfnamefont {S.}~\bibnamefont {{De Liberato}}}, \ and\
  \bibinfo {author} {\bibfnamefont {M.}~\bibnamefont {Ueda}},\ }\bibfield
  {title} {\enquote {\bibinfo {title} {{Quantum Szilard Engine}},}\ }\href
  {\doibase 10.1103/PhysRevLett.106.070401} {\bibfield  {journal} {\bibinfo
  {journal} {PRL}\ }\textbf {\bibinfo {volume} {106}},\ \bibinfo {pages}
  {70401} (\bibinfo {year} {2011})}\BibitemShut {NoStop}%
\bibitem [{\citenamefont {Erker}\ \emph {et~al.}(2017)\citenamefont {Erker},
  \citenamefont {Mitchison}, \citenamefont {Silva}, \citenamefont {Woods},
  \citenamefont {Brunner},\ and\ \citenamefont {Huber}}]{Erker2017}%
  \BibitemOpen
  \bibfield  {author} {\bibinfo {author} {\bibfnamefont {P.}~\bibnamefont
  {Erker}}, \bibinfo {author} {\bibfnamefont {M.~T.}\ \bibnamefont
  {Mitchison}}, \bibinfo {author} {\bibfnamefont {R.}~\bibnamefont {Silva}},
  \bibinfo {author} {\bibfnamefont {M.~P.}\ \bibnamefont {Woods}}, \bibinfo
  {author} {\bibfnamefont {N.}~\bibnamefont {Brunner}}, \ and\ \bibinfo
  {author} {\bibfnamefont {M.}~\bibnamefont {Huber}},\ }\bibfield  {title}
  {\enquote {\bibinfo {title} {{Autonomous Quantum Clocks: Does Thermodynamics
  Limit Our Ability to Measure Time?}}}\ }\href {\doibase
  10.1103/PhysRevX.7.031022} {\bibfield  {journal} {\bibinfo  {journal} {Phys.
  Rev. X}\ }\textbf {\bibinfo {volume} {7}},\ \bibinfo {pages} {031022}
  (\bibinfo {year} {2017})},\ \Eprint {http://arxiv.org/abs/1609.06704}
  {arXiv:1609.06704} \BibitemShut {NoStop}%
\bibitem [{\citenamefont {Pearson}\ \emph {et~al.}(2021)\citenamefont
  {Pearson}, \citenamefont {Guryanova}, \citenamefont {Erker}, \citenamefont
  {Laird}, \citenamefont {Briggs}, \citenamefont {Huber},\ and\ \citenamefont
  {Ares}}]{Pearson2021}%
  \BibitemOpen
  \bibfield  {author} {\bibinfo {author} {\bibfnamefont {A.~N.}\ \bibnamefont
  {Pearson}}, \bibinfo {author} {\bibfnamefont {Y.}~\bibnamefont {Guryanova}},
  \bibinfo {author} {\bibfnamefont {P.}~\bibnamefont {Erker}}, \bibinfo
  {author} {\bibfnamefont {E.~A.}\ \bibnamefont {Laird}}, \bibinfo {author}
  {\bibfnamefont {G.~A.~D.}\ \bibnamefont {Briggs}}, \bibinfo {author}
  {\bibfnamefont {M.}~\bibnamefont {Huber}}, \ and\ \bibinfo {author}
  {\bibfnamefont {N.}~\bibnamefont {Ares}},\ }\bibfield  {title} {\enquote
  {\bibinfo {title} {{Measuring the Thermodynamic Cost of Timekeeping}},}\
  }\href {\doibase 10.1103/PhysRevX.11.021029} {\bibfield  {journal} {\bibinfo
  {journal} {Phys. Rev. X}\ }\textbf {\bibinfo {volume} {11}},\ \bibinfo
  {pages} {021029} (\bibinfo {year} {2021})},\ \Eprint
  {http://arxiv.org/abs/2006.08670} {arXiv:2006.08670} \BibitemShut {NoStop}%
\bibitem [{\citenamefont {Yang}\ and\ \citenamefont
  {Halperin}(2009)}]{Yang2009}%
  \BibitemOpen
  \bibfield  {author} {\bibinfo {author} {\bibfnamefont {K.}~\bibnamefont
  {Yang}}\ and\ \bibinfo {author} {\bibfnamefont {B.~I.}\ \bibnamefont
  {Halperin}},\ }\bibfield  {title} {\enquote {\bibinfo {title} {{Thermopower
  as a possible probe of non-Abelian quasiparticle statistics in fractional
  quantum Hall liquids}},}\ }\href {\doibase 10.1103/PhysRevB.79.115317}
  {\bibfield  {journal} {\bibinfo  {journal} {Phys. Rev. B - Condens. Matter
  Mater. Phys.}\ }\textbf {\bibinfo {volume} {79}},\ \bibinfo {pages} {1--5}
  (\bibinfo {year} {2009})},\ \Eprint {http://arxiv.org/abs/0901.1429}
  {arXiv:0901.1429} \BibitemShut {NoStop}%
\bibitem [{\citenamefont {Chaikin}\ and\ \citenamefont
  {Beni}(1976)}]{Chaikin1976}%
  \BibitemOpen
  \bibfield  {author} {\bibinfo {author} {\bibfnamefont {P.~M.}\ \bibnamefont
  {Chaikin}}\ and\ \bibinfo {author} {\bibfnamefont {G.}~\bibnamefont {Beni}},\
  }\bibfield  {title} {\enquote {\bibinfo {title} {{Thermopower in the
  correlated hopping regime}},}\ }\href {\doibase 10.1103/PhysRevB.13.647}
  {\bibfield  {journal} {\bibinfo  {journal} {Phys. Rev. B}\ }\textbf {\bibinfo
  {volume} {13}},\ \bibinfo {pages} {647--651} (\bibinfo {year}
  {1976})}\BibitemShut {NoStop}%
\bibitem [{\citenamefont {Goupil}\ \emph {et~al.}(2016)\citenamefont {Goupil},
  \citenamefont {Ouerdane}, \citenamefont {Zabrocki}, \citenamefont {Seifert},
  \citenamefont {Hinsche},\ and\ \citenamefont {M{\"{u}}ller}}]{Goupil2016}%
  \BibitemOpen
  \bibfield  {author} {\bibinfo {author} {\bibfnamefont {C.}~\bibnamefont
  {Goupil}}, \bibinfo {author} {\bibfnamefont {H.}~\bibnamefont {Ouerdane}},
  \bibinfo {author} {\bibfnamefont {K.}~\bibnamefont {Zabrocki}}, \bibinfo
  {author} {\bibfnamefont {W.}~\bibnamefont {Seifert}}, \bibinfo {author}
  {\bibfnamefont {N.~F.}\ \bibnamefont {Hinsche}}, \ and\ \bibinfo {author}
  {\bibfnamefont {E.}~\bibnamefont {M{\"{u}}ller}},\ }\bibfield  {title}
  {\enquote {\bibinfo {title} {{Thermodynamics and Thermoelectricity}},}\ }in\
  \href {\doibase 10.1002/9783527338405.ch1} {\emph {\bibinfo {booktitle}
  {Contin. Theory andModeling Thermoelectr. Elem.}}}\ (\bibinfo  {publisher}
  {Wiley},\ \bibinfo {year} {2016})\ pp.\ \bibinfo {pages} {1--74}\BibitemShut
  {NoStop}%
\bibitem [{\citenamefont {Behnia}, \citenamefont {Jaccard},\ and\ \citenamefont
  {Flouquet}(2004)}]{Behnia2004}%
  \BibitemOpen
  \bibfield  {author} {\bibinfo {author} {\bibfnamefont {K.}~\bibnamefont
  {Behnia}}, \bibinfo {author} {\bibfnamefont {D.}~\bibnamefont {Jaccard}}, \
  and\ \bibinfo {author} {\bibfnamefont {J.}~\bibnamefont {Flouquet}},\
  }\bibfield  {title} {\enquote {\bibinfo {title} {{On the thermoelectricity of
  correlated electrons in the zero-temperature limit}},}\ }\href {\doibase
  10.1088/0953-8984/16/28/037} {\bibfield  {journal} {\bibinfo  {journal} {J.
  Phys. Condens. Matter}\ }\textbf {\bibinfo {volume} {16}},\ \bibinfo {pages}
  {5187--5198} (\bibinfo {year} {2004})},\ \Eprint
  {http://arxiv.org/abs/0405030} {arXiv:0405030 [cond-mat]} \BibitemShut
  {NoStop}%
\bibitem [{\citenamefont {Zlati{\'{c}}}\ \emph {et~al.}(2007)\citenamefont
  {Zlati{\'{c}}}, \citenamefont {Monnier}, \citenamefont {Freericks},\ and\
  \citenamefont {Becker}}]{Zlatic2007}%
  \BibitemOpen
  \bibfield  {author} {\bibinfo {author} {\bibfnamefont {V.}~\bibnamefont
  {Zlati{\'{c}}}}, \bibinfo {author} {\bibfnamefont {R.}~\bibnamefont
  {Monnier}}, \bibinfo {author} {\bibfnamefont {J.~K.}\ \bibnamefont
  {Freericks}}, \ and\ \bibinfo {author} {\bibfnamefont {K.~W.}\ \bibnamefont
  {Becker}},\ }\bibfield  {title} {\enquote {\bibinfo {title} {{Relationship
  between the thermopower and entropy of strongly correlated electron
  systems}},}\ }\href {\doibase 10.1103/PhysRevB.76.085122} {\bibfield
  {journal} {\bibinfo  {journal} {Phys. Rev. B - Condens. Matter Mater. Phys.}\
  }\textbf {\bibinfo {volume} {76}},\ \bibinfo {pages} {1--16} (\bibinfo {year}
  {2007})},\ \Eprint {http://arxiv.org/abs/0512288} {arXiv:0512288 [cond-mat]}
  \BibitemShut {NoStop}%
\bibitem [{\citenamefont {Mravlje}\ and\ \citenamefont
  {Georges}(2016)}]{Mravlje2016}%
  \BibitemOpen
  \bibfield  {author} {\bibinfo {author} {\bibfnamefont {J.}~\bibnamefont
  {Mravlje}}\ and\ \bibinfo {author} {\bibfnamefont {A.}~\bibnamefont
  {Georges}},\ }\bibfield  {title} {\enquote {\bibinfo {title} {{Thermopower
  and Entropy: Lessons from Sr$_2$RuO$_4$}},}\ }\href {\doibase
  10.1103/PhysRevLett.117.036401} {\bibfield  {journal} {\bibinfo  {journal}
  {Phys. Rev. Lett.}\ }\textbf {\bibinfo {volume} {117}},\ \bibinfo {pages}
  {036401} (\bibinfo {year} {2016})},\ \Eprint
  {http://arxiv.org/abs/1504.03860} {arXiv:1504.03860} \BibitemShut {NoStop}%
\bibitem [{\citenamefont {Cooper}\ and\ \citenamefont
  {Stern}(2009)}]{Cooper2009}%
  \BibitemOpen
  \bibfield  {author} {\bibinfo {author} {\bibfnamefont {N.~R.}\ \bibnamefont
  {Cooper}}\ and\ \bibinfo {author} {\bibfnamefont {A.}~\bibnamefont {Stern}},\
  }\bibfield  {title} {\enquote {\bibinfo {title} {{Observable bulk signatures
  of Non-Abelian quantum hall states}},}\ }\href {\doibase
  10.1103/PhysRevLett.102.176807} {\bibfield  {journal} {\bibinfo  {journal}
  {Phys. Rev. Lett.}\ }\textbf {\bibinfo {volume} {102}},\ \bibinfo {pages}
  {1--4} (\bibinfo {year} {2009})},\ \Eprint {http://arxiv.org/abs/0812.3387}
  {arXiv:0812.3387} \BibitemShut {NoStop}%
\bibitem [{\citenamefont {Sarma}, \citenamefont {Freedman},\ and\ \citenamefont
  {Nayak}(2005)}]{Sarma2005}%
  \BibitemOpen
  \bibfield  {author} {\bibinfo {author} {\bibfnamefont {S.~D.}\ \bibnamefont
  {Sarma}}, \bibinfo {author} {\bibfnamefont {M.}~\bibnamefont {Freedman}}, \
  and\ \bibinfo {author} {\bibfnamefont {C.}~\bibnamefont {Nayak}},\ }\bibfield
   {title} {\enquote {\bibinfo {title} {{Topologically protected qubits from a
  possible non-abelian fractional quantum hall state}},}\ }\href {\doibase
  10.1103/PhysRevLett.94.166802} {\bibfield  {journal} {\bibinfo  {journal}
  {Phys. Rev. Lett.}\ }\textbf {\bibinfo {volume} {94}},\ \bibinfo {pages}
  {1--4} (\bibinfo {year} {2005})},\ \Eprint {http://arxiv.org/abs/0412343}
  {arXiv:0412343 [cond-mat]} \BibitemShut {NoStop}%
\bibitem [{\citenamefont {Stern}\ and\ \citenamefont
  {Halperin}(2006)}]{Stern2006}%
  \BibitemOpen
  \bibfield  {author} {\bibinfo {author} {\bibfnamefont {A.}~\bibnamefont
  {Stern}}\ and\ \bibinfo {author} {\bibfnamefont {B.~I.}\ \bibnamefont
  {Halperin}},\ }\bibfield  {title} {\enquote {\bibinfo {title} {{Proposed
  Experiments to Probe the Non-Abelian $\nu$=5/2 Quantum Hall State}},}\ }\href
  {\doibase 10.1103/PhysRevLett.96.016802} {\bibfield  {journal} {\bibinfo
  {journal} {Phys. Rev. Lett.}\ }\textbf {\bibinfo {volume} {96}},\ \bibinfo
  {pages} {016802} (\bibinfo {year} {2006})},\ \Eprint
  {http://arxiv.org/abs/0508447} {arXiv:0508447 [cond-mat]} \BibitemShut
  {NoStop}%
\bibitem [{\citenamefont {Bonderson}, \citenamefont {Kitaev},\ and\
  \citenamefont {Shtengel}(2006)}]{Bonderson2006}%
  \BibitemOpen
  \bibfield  {author} {\bibinfo {author} {\bibfnamefont {P.}~\bibnamefont
  {Bonderson}}, \bibinfo {author} {\bibfnamefont {A.}~\bibnamefont {Kitaev}}, \
  and\ \bibinfo {author} {\bibfnamefont {K.}~\bibnamefont {Shtengel}},\
  }\bibfield  {title} {\enquote {\bibinfo {title} {{Detecting Non-Abelian
  Statistics in the $\nu$=5/2 Fractional Quantum Hall State}},}\ }\href
  {\doibase 10.1103/PhysRevLett.96.016803} {\bibfield  {journal} {\bibinfo
  {journal} {Phys. Rev. Lett.}\ }\textbf {\bibinfo {volume} {96}},\ \bibinfo
  {pages} {016803} (\bibinfo {year} {2006})},\ \Eprint
  {http://arxiv.org/abs/0508616} {arXiv:0508616 [cond-mat]} \BibitemShut
  {NoStop}%
\bibitem [{\citenamefont {Feldman}\ \emph {et~al.}(2007)\citenamefont
  {Feldman}, \citenamefont {Gefen}, \citenamefont {Kitaev}, \citenamefont
  {Law},\ and\ \citenamefont {Stern}}]{Feldman2007}%
  \BibitemOpen
  \bibfield  {author} {\bibinfo {author} {\bibfnamefont {D.~E.}\ \bibnamefont
  {Feldman}}, \bibinfo {author} {\bibfnamefont {Y.}~\bibnamefont {Gefen}},
  \bibinfo {author} {\bibfnamefont {A.}~\bibnamefont {Kitaev}}, \bibinfo
  {author} {\bibfnamefont {K.~T.}\ \bibnamefont {Law}}, \ and\ \bibinfo
  {author} {\bibfnamefont {A.}~\bibnamefont {Stern}},\ }\bibfield  {title}
  {\enquote {\bibinfo {title} {{Shot noise in an anyonic Mach-Zehnder
  interferometer}},}\ }\href {\doibase 10.1103/PhysRevB.76.085333} {\bibfield
  {journal} {\bibinfo  {journal} {Phys. Rev. B - Condens. Matter Mater. Phys.}\
  }\textbf {\bibinfo {volume} {76}},\ \bibinfo {pages} {1--9} (\bibinfo {year}
  {2007})},\ \Eprint {http://arxiv.org/abs/0612608} {arXiv:0612608 [cond-mat]}
  \BibitemShut {NoStop}%
\bibitem [{\citenamefont {Hofmann}\ \emph {et~al.}(2016)\citenamefont
  {Hofmann}, \citenamefont {Maisi}, \citenamefont {Gold}, \citenamefont
  {Kr{\"{a}}henmann}, \citenamefont {R{\"{o}}ssler}, \citenamefont {Basset},
  \citenamefont {M{\"{a}}rki}, \citenamefont {Reichl}, \citenamefont
  {Wegscheider}, \citenamefont {Ensslin},\ and\ \citenamefont
  {Ihn}}]{Hofmann2016}%
  \BibitemOpen
  \bibfield  {author} {\bibinfo {author} {\bibfnamefont {A.}~\bibnamefont
  {Hofmann}}, \bibinfo {author} {\bibfnamefont {V.~F.}\ \bibnamefont {Maisi}},
  \bibinfo {author} {\bibfnamefont {C.}~\bibnamefont {Gold}}, \bibinfo {author}
  {\bibfnamefont {T.}~\bibnamefont {Kr{\"{a}}henmann}}, \bibinfo {author}
  {\bibfnamefont {C.}~\bibnamefont {R{\"{o}}ssler}}, \bibinfo {author}
  {\bibfnamefont {J.}~\bibnamefont {Basset}}, \bibinfo {author} {\bibfnamefont
  {P.}~\bibnamefont {M{\"{a}}rki}}, \bibinfo {author} {\bibfnamefont
  {C.}~\bibnamefont {Reichl}}, \bibinfo {author} {\bibfnamefont
  {W.}~\bibnamefont {Wegscheider}}, \bibinfo {author} {\bibfnamefont
  {K.}~\bibnamefont {Ensslin}}, \ and\ \bibinfo {author} {\bibfnamefont
  {T.}~\bibnamefont {Ihn}},\ }\bibfield  {title} {\enquote {\bibinfo {title}
  {{Measuring the degeneracy of discrete energy levels using a GaAs/AlGaAs
  quantum dot}},}\ }\href {\doibase 10.1103/PhysRevLett.117.206803} {\bibfield
  {journal} {\bibinfo  {journal} {Phys. Rev. Lett.}\ }\textbf {\bibinfo
  {volume} {117}},\ \bibinfo {pages} {1--6} (\bibinfo {year} {2016})},\ \Eprint
  {http://arxiv.org/abs/1610.00928} {arXiv:1610.00928} \BibitemShut {NoStop}%
\bibitem [{\citenamefont {Harzheim}\ \emph {et~al.}(2020)\citenamefont
  {Harzheim}, \citenamefont {Sowa}, \citenamefont {Swett}, \citenamefont
  {Briggs}, \citenamefont {Mol},\ and\ \citenamefont {Gehring}}]{Harzheim2020}%
  \BibitemOpen
  \bibfield  {author} {\bibinfo {author} {\bibfnamefont {A.}~\bibnamefont
  {Harzheim}}, \bibinfo {author} {\bibfnamefont {J.~K.}\ \bibnamefont {Sowa}},
  \bibinfo {author} {\bibfnamefont {J.~L.}\ \bibnamefont {Swett}}, \bibinfo
  {author} {\bibfnamefont {G.~A.~D.}\ \bibnamefont {Briggs}}, \bibinfo {author}
  {\bibfnamefont {J.~A.}\ \bibnamefont {Mol}}, \ and\ \bibinfo {author}
  {\bibfnamefont {P.}~\bibnamefont {Gehring}},\ }\bibfield  {title} {\enquote
  {\bibinfo {title} {{Role of metallic leads and electronic degeneracies in
  thermoelectric power generation in quantum dots}},}\ }\href {\doibase
  10.1103/PhysRevResearch.2.013140} {\bibfield  {journal} {\bibinfo  {journal}
  {Phys. Rev. Res.}\ }\textbf {\bibinfo {volume} {2}},\ \bibinfo {pages}
  {013140} (\bibinfo {year} {2020})},\ \Eprint
  {http://arxiv.org/abs/1906.05401} {arXiv:1906.05401} \BibitemShut {NoStop}%
\bibitem [{\citenamefont {Nazarov}\ and\ \citenamefont
  {Blanter}(2009)}]{Nazarov2009}%
  \BibitemOpen
  \bibfield  {author} {\bibinfo {author} {\bibfnamefont {Y.~V.}\ \bibnamefont
  {Nazarov}}\ and\ \bibinfo {author} {\bibfnamefont {Y.~M.}\ \bibnamefont
  {Blanter}},\ }\href {\doibase 10.1017/CBO9780511626906} {\emph {\bibinfo
  {title} {{Quantum Transport}}}}\ (\bibinfo  {publisher} {Cambridge University
  Press},\ \bibinfo {address} {Cambridge},\ \bibinfo {year} {2009})\BibitemShut
  {NoStop}%
\bibitem [{\citenamefont {Hanson}\ \emph {et~al.}(2007)\citenamefont {Hanson},
  \citenamefont {Kouwenhoven}, \citenamefont {Petta}, \citenamefont {Tarucha},\
  and\ \citenamefont {Vandersypen}}]{Hanson2007}%
  \BibitemOpen
  \bibfield  {author} {\bibinfo {author} {\bibfnamefont {R.}~\bibnamefont
  {Hanson}}, \bibinfo {author} {\bibfnamefont {L.~P.}\ \bibnamefont
  {Kouwenhoven}}, \bibinfo {author} {\bibfnamefont {J.~R.}\ \bibnamefont
  {Petta}}, \bibinfo {author} {\bibfnamefont {S.}~\bibnamefont {Tarucha}}, \
  and\ \bibinfo {author} {\bibfnamefont {L.~M.}\ \bibnamefont {Vandersypen}},\
  }\bibfield  {title} {\enquote {\bibinfo {title} {{Spins in few-electron
  quantum dots}},}\ }\href {\doibase 10.1103/RevModPhys.79.1217} {\bibfield
  {journal} {\bibinfo  {journal} {Rev. Mod. Phys.}\ }\textbf {\bibinfo {volume}
  {79}},\ \bibinfo {pages} {1217--1265} (\bibinfo {year} {2007})},\ \Eprint
  {http://arxiv.org/abs/0610433} {arXiv:0610433 [cond-mat]} \BibitemShut
  {NoStop}%
\bibitem [{\citenamefont {Thijssen}\ and\ \citenamefont {{Van der
  Zant}}(2008)}]{Thijssen2008}%
  \BibitemOpen
  \bibfield  {author} {\bibinfo {author} {\bibfnamefont {J.~M.}\ \bibnamefont
  {Thijssen}}\ and\ \bibinfo {author} {\bibfnamefont {H.~S.~J.}\ \bibnamefont
  {{Van der Zant}}},\ }\bibfield  {title} {\enquote {\bibinfo {title} {{Charge
  transport and single-electron effects in nanoscale systems}},}\ }\href
  {\doibase 10.1002/pssb.200743470} {\bibfield  {journal} {\bibinfo  {journal}
  {Phys. status solidi}\ }\textbf {\bibinfo {volume} {245}},\ \bibinfo {pages}
  {1455--1470} (\bibinfo {year} {2008})}\BibitemShut {NoStop}%
\bibitem [{\citenamefont {Fuechsle}\ \emph {et~al.}(2012)\citenamefont
  {Fuechsle}, \citenamefont {Miwa}, \citenamefont {Mahapatra}, \citenamefont
  {Ryu}, \citenamefont {Lee}, \citenamefont {Warschkow}, \citenamefont
  {Hollenberg}, \citenamefont {Klimeck},\ and\ \citenamefont
  {Simmons}}]{Fuechsle2012}%
  \BibitemOpen
  \bibfield  {author} {\bibinfo {author} {\bibfnamefont {M.}~\bibnamefont
  {Fuechsle}}, \bibinfo {author} {\bibfnamefont {J.~A.}\ \bibnamefont {Miwa}},
  \bibinfo {author} {\bibfnamefont {S.}~\bibnamefont {Mahapatra}}, \bibinfo
  {author} {\bibfnamefont {H.}~\bibnamefont {Ryu}}, \bibinfo {author}
  {\bibfnamefont {S.}~\bibnamefont {Lee}}, \bibinfo {author} {\bibfnamefont
  {O.}~\bibnamefont {Warschkow}}, \bibinfo {author} {\bibfnamefont {L.~C.~L.}\
  \bibnamefont {Hollenberg}}, \bibinfo {author} {\bibfnamefont
  {G.}~\bibnamefont {Klimeck}}, \ and\ \bibinfo {author} {\bibfnamefont
  {M.~Y.}\ \bibnamefont {Simmons}},\ }\bibfield  {title} {\enquote {\bibinfo
  {title} {A single-atom transistor},}\ }\href {\doibase 10.1038/nnano.2012.21}
  {\bibfield  {journal} {\bibinfo  {journal} {Nature Nanotechnology}\ }\textbf
  {\bibinfo {volume} {7}},\ \bibinfo {pages} {242--246} (\bibinfo {year}
  {2012})}\BibitemShut {NoStop}%
\bibitem [{\citenamefont {Osorio}\ \emph {et~al.}(2008)\citenamefont {Osorio},
  \citenamefont {Bj{\o}rnholm}, \citenamefont {Lehn}, \citenamefont {Ruben},\
  and\ \citenamefont {{Van Der Zant}}}]{Osorio2008}%
  \BibitemOpen
  \bibfield  {author} {\bibinfo {author} {\bibfnamefont {E.~A.}\ \bibnamefont
  {Osorio}}, \bibinfo {author} {\bibfnamefont {T.}~\bibnamefont
  {Bj{\o}rnholm}}, \bibinfo {author} {\bibfnamefont {J.~M.}\ \bibnamefont
  {Lehn}}, \bibinfo {author} {\bibfnamefont {M.}~\bibnamefont {Ruben}}, \ and\
  \bibinfo {author} {\bibfnamefont {H.~S.}\ \bibnamefont {{Van Der Zant}}},\
  }\bibfield  {title} {\enquote {\bibinfo {title} {{Single-molecule transport
  in three-terminal devices}},}\ }\href {\doibase
  10.1088/0953-8984/20/37/374121} {\bibfield  {journal} {\bibinfo  {journal}
  {J. Phys. Condens. Matter}\ }\textbf {\bibinfo {volume} {20}} (\bibinfo
  {year} {2008}),\ 10.1088/0953-8984/20/37/374121}\BibitemShut {NoStop}%
\bibitem [{\citenamefont {Brooke}\ \emph {et~al.}(2015)\citenamefont {Brooke},
  \citenamefont {Vezzoli}, \citenamefont {Higgins}, \citenamefont {Zotti},
  \citenamefont {Palacios},\ and\ \citenamefont {Nichols}}]{Brooke2015}%
  \BibitemOpen
  \bibfield  {author} {\bibinfo {author} {\bibfnamefont {C.}~\bibnamefont
  {Brooke}}, \bibinfo {author} {\bibfnamefont {A.}~\bibnamefont {Vezzoli}},
  \bibinfo {author} {\bibfnamefont {S.~J.}\ \bibnamefont {Higgins}}, \bibinfo
  {author} {\bibfnamefont {L.~A.}\ \bibnamefont {Zotti}}, \bibinfo {author}
  {\bibfnamefont {J.~J.}\ \bibnamefont {Palacios}}, \ and\ \bibinfo {author}
  {\bibfnamefont {R.~J.}\ \bibnamefont {Nichols}},\ }\bibfield  {title}
  {\enquote {\bibinfo {title} {{Resonant transport and electrostatic effects in
  single-molecule electrical junctions}},}\ }\href {\doibase
  10.1103/PhysRevB.91.195438} {\bibfield  {journal} {\bibinfo  {journal} {Phys.
  Rev. B - Condens. Matter Mater. Phys.}\ }\textbf {\bibinfo {volume} {91}},\
  \bibinfo {pages} {1--9} (\bibinfo {year} {2015})}\BibitemShut {NoStop}%
\bibitem [{\citenamefont {Blundell}\ and\ \citenamefont
  {Blundell}(2009)}]{Blundell}%
  \BibitemOpen
  \bibfield  {author} {\bibinfo {author} {\bibfnamefont {S.~J.}\ \bibnamefont
  {Blundell}}\ and\ \bibinfo {author} {\bibfnamefont {K.~M.}\ \bibnamefont
  {Blundell}},\ }\href {\doibase 10.1093/acprof:oso/9780199562091.001.0001}
  {\emph {\bibinfo {title} {{Concepts in Thermal Physics}}}}\ (\bibinfo
  {publisher} {Oxford University Press},\ \bibinfo {year} {2009})\BibitemShut
  {NoStop}%
\bibitem [{\citenamefont {Ford}(2013)}]{Ford2013}%
  \BibitemOpen
  \bibfield  {author} {\bibinfo {author} {\bibfnamefont {I.}~\bibnamefont
  {Ford}},\ }\href {\doibase 10.1002/9781118597507} {\emph {\bibinfo {title}
  {{Statistical Physics: An Entropic Approach}}}}\ (\bibinfo  {publisher} {John
  Wiley {\&} Sons, Ltd},\ \bibinfo {address} {Chichester, UK},\ \bibinfo {year}
  {2013})\BibitemShut {NoStop}%
\bibitem [{\citenamefont {Child}\ \emph
  {et~al.}(2021{\natexlab{a}})\citenamefont {Child}, \citenamefont {Sheekey},
  \citenamefont {L{\"{u}}scher}, \citenamefont {Fallahi}, \citenamefont
  {Gardner}, \citenamefont {Manfra}, \citenamefont {Kleeorin}, \citenamefont
  {Meir},\ and\ \citenamefont {Folk}}]{Child2021a}%
  \BibitemOpen
  \bibfield  {author} {\bibinfo {author} {\bibfnamefont {T.}~\bibnamefont
  {Child}}, \bibinfo {author} {\bibfnamefont {O.}~\bibnamefont {Sheekey}},
  \bibinfo {author} {\bibfnamefont {S.}~\bibnamefont {L{\"{u}}scher}}, \bibinfo
  {author} {\bibfnamefont {S.}~\bibnamefont {Fallahi}}, \bibinfo {author}
  {\bibfnamefont {G.~C.}\ \bibnamefont {Gardner}}, \bibinfo {author}
  {\bibfnamefont {M.}~\bibnamefont {Manfra}}, \bibinfo {author} {\bibfnamefont
  {Y.}~\bibnamefont {Kleeorin}}, \bibinfo {author} {\bibfnamefont
  {Y.}~\bibnamefont {Meir}}, \ and\ \bibinfo {author} {\bibfnamefont
  {J.}~\bibnamefont {Folk}},\ }\bibfield  {title} {\enquote {\bibinfo {title}
  {{Entropy measurement of a strongly correlated quantum dot}},}\ }\href
  {http://arxiv.org/abs/2110.14158} {\  (\bibinfo {year}
  {2021}{\natexlab{a}})},\ \Eprint {http://arxiv.org/abs/2110.14158}
  {arXiv:2110.14158} \BibitemShut {NoStop}%
\bibitem [{\citenamefont {Sela}\ \emph {et~al.}(2019)\citenamefont {Sela},
  \citenamefont {Oreg}, \citenamefont {Plugge}, \citenamefont {Hartman},
  \citenamefont {L\"uscher},\ and\ \citenamefont {Folk}}]{Sela2019}%
  \BibitemOpen
  \bibfield  {author} {\bibinfo {author} {\bibfnamefont {E.}~\bibnamefont
  {Sela}}, \bibinfo {author} {\bibfnamefont {Y.}~\bibnamefont {Oreg}}, \bibinfo
  {author} {\bibfnamefont {S.}~\bibnamefont {Plugge}}, \bibinfo {author}
  {\bibfnamefont {N.}~\bibnamefont {Hartman}}, \bibinfo {author} {\bibfnamefont
  {S.}~\bibnamefont {L\"uscher}}, \ and\ \bibinfo {author} {\bibfnamefont
  {J.}~\bibnamefont {Folk}},\ }\bibfield  {title} {\enquote {\bibinfo {title}
  {Detecting the universal fractional entropy of majorana zero modes},}\ }\href
  {\doibase 10.1103/PhysRevLett.123.147702} {\bibfield  {journal} {\bibinfo
  {journal} {Phys. Rev. Lett.}\ }\textbf {\bibinfo {volume} {123}},\ \bibinfo
  {pages} {147702} (\bibinfo {year} {2019})}\BibitemShut {NoStop}%
\bibitem [{\citenamefont {Child}\ \emph
  {et~al.}(2021{\natexlab{b}})\citenamefont {Child}, \citenamefont {Sheekey},
  \citenamefont {L{\"{u}}scher}, \citenamefont {Fallahi}, \citenamefont
  {Gardner}, \citenamefont {Manfra},\ and\ \citenamefont {Folk}}]{Child2021}%
  \BibitemOpen
  \bibfield  {author} {\bibinfo {author} {\bibfnamefont {T.}~\bibnamefont
  {Child}}, \bibinfo {author} {\bibfnamefont {O.}~\bibnamefont {Sheekey}},
  \bibinfo {author} {\bibfnamefont {S.}~\bibnamefont {L{\"{u}}scher}}, \bibinfo
  {author} {\bibfnamefont {S.}~\bibnamefont {Fallahi}}, \bibinfo {author}
  {\bibfnamefont {G.~C.}\ \bibnamefont {Gardner}}, \bibinfo {author}
  {\bibfnamefont {M.}~\bibnamefont {Manfra}}, \ and\ \bibinfo {author}
  {\bibfnamefont {J.}~\bibnamefont {Folk}},\ }\bibfield  {title} {\enquote
  {\bibinfo {title} {{A robust protocol for entropy measurement in mesoscopic
  circuits}},}\ }\href {http://arxiv.org/abs/2110.14172} {\ ,\ \bibinfo {pages}
  {1--7} (\bibinfo {year} {2021}{\natexlab{b}})},\ \Eprint
  {http://arxiv.org/abs/2110.14172} {arXiv:2110.14172} \BibitemShut {NoStop}%
\bibitem [{\citenamefont {Pyurbeeva}\ \emph {et~al.}(2021)\citenamefont
  {Pyurbeeva}, \citenamefont {Hsu}, \citenamefont {Vogel}, \citenamefont
  {Wegeberg}, \citenamefont {Mayor}, \citenamefont {{Van Der Zant}},
  \citenamefont {Mol},\ and\ \citenamefont {Gehring}}]{Pyurbeeva2021b}%
  \BibitemOpen
  \bibfield  {author} {\bibinfo {author} {\bibfnamefont {E.}~\bibnamefont
  {Pyurbeeva}}, \bibinfo {author} {\bibfnamefont {C.}~\bibnamefont {Hsu}},
  \bibinfo {author} {\bibfnamefont {D.}~\bibnamefont {Vogel}}, \bibinfo
  {author} {\bibfnamefont {C.}~\bibnamefont {Wegeberg}}, \bibinfo {author}
  {\bibfnamefont {M.}~\bibnamefont {Mayor}}, \bibinfo {author} {\bibfnamefont
  {H.}~\bibnamefont {{Van Der Zant}}}, \bibinfo {author} {\bibfnamefont
  {J.~A.}\ \bibnamefont {Mol}}, \ and\ \bibinfo {author} {\bibfnamefont
  {P.}~\bibnamefont {Gehring}},\ }\bibfield  {title} {\enquote {\bibinfo
  {title} {{Controlling the Entropy of a Single-Molecule Junction}},}\ }\href
  {\doibase 10.1021/acs.nanolett.1c03591} {\bibfield  {journal} {\bibinfo
  {journal} {Nano Lett.}\ }\textbf {\bibinfo {volume} {21}},\ \bibinfo {pages}
  {9715--9719} (\bibinfo {year} {2021})},\ \Eprint
  {http://arxiv.org/abs/2109.06741} {arXiv:2109.06741} \BibitemShut {NoStop}%
\bibitem [{\citenamefont {Seldenthuis}\ \emph {et~al.}(2008)\citenamefont
  {Seldenthuis}, \citenamefont {van~der Zant}, \citenamefont {Ratner},\ and\
  \citenamefont {Thijssen}}]{Seldenthuis08}%
  \BibitemOpen
  \bibfield  {author} {\bibinfo {author} {\bibfnamefont {J.~S.}\ \bibnamefont
  {Seldenthuis}}, \bibinfo {author} {\bibfnamefont {H.~S.~J.}\ \bibnamefont
  {van~der Zant}}, \bibinfo {author} {\bibfnamefont {M.~A.}\ \bibnamefont
  {Ratner}}, \ and\ \bibinfo {author} {\bibfnamefont {J.~M.}\ \bibnamefont
  {Thijssen}},\ }\bibfield  {title} {\enquote {\bibinfo {title} {Vibrational
  excitations in weakly coupled single-molecule junctions: A computational
  analysis},}\ }\href {\doibase 10.1021/nn800170h} {\bibfield  {journal}
  {\bibinfo  {journal} {ACS Nano}\ }\textbf {\bibinfo {volume} {2}},\ \bibinfo
  {pages} {1445--1451} (\bibinfo {year} {2008})},\ \Eprint
  {http://arxiv.org/abs/https://doi.org/10.1021/nn800170h}
  {https://doi.org/10.1021/nn800170h} \BibitemShut {NoStop}%
\bibitem [{\citenamefont {Thomas}\ \emph {et~al.}(2019)\citenamefont {Thomas},
  \citenamefont {Limburg}, \citenamefont {Sowa}, \citenamefont {Willick},
  \citenamefont {Baugh}, \citenamefont {Briggs}, \citenamefont {Gauger},
  \citenamefont {Anderson},\ and\ \citenamefont {Mol}}]{Thomas2019}%
  \BibitemOpen
  \bibfield  {author} {\bibinfo {author} {\bibfnamefont {J.~O.}\ \bibnamefont
  {Thomas}}, \bibinfo {author} {\bibfnamefont {B.}~\bibnamefont {Limburg}},
  \bibinfo {author} {\bibfnamefont {J.~K.}\ \bibnamefont {Sowa}}, \bibinfo
  {author} {\bibfnamefont {K.}~\bibnamefont {Willick}}, \bibinfo {author}
  {\bibfnamefont {J.}~\bibnamefont {Baugh}}, \bibinfo {author} {\bibfnamefont
  {G.~A.~D.}\ \bibnamefont {Briggs}}, \bibinfo {author} {\bibfnamefont {E.~M.}\
  \bibnamefont {Gauger}}, \bibinfo {author} {\bibfnamefont {H.~L.}\
  \bibnamefont {Anderson}}, \ and\ \bibinfo {author} {\bibfnamefont {J.~A.}\
  \bibnamefont {Mol}},\ }\bibfield  {title} {\enquote {\bibinfo {title}
  {Understanding resonant charge transport through weakly coupled
  single-molecule junctions},}\ }\href {\doibase 10.1038/s41467-019-12625-4}
  {\bibfield  {journal} {\bibinfo  {journal} {Nature Communications}\ }\textbf
  {\bibinfo {volume} {10}},\ \bibinfo {pages} {4628} (\bibinfo {year}
  {2019})}\BibitemShut {NoStop}%
\bibitem [{\citenamefont {Sowa}, \citenamefont {Mol},\ and\ \citenamefont
  {Gauger}(2019)}]{Sowa2019}%
  \BibitemOpen
  \bibfield  {author} {\bibinfo {author} {\bibfnamefont {J.~K.}\ \bibnamefont
  {Sowa}}, \bibinfo {author} {\bibfnamefont {J.~A.}\ \bibnamefont {Mol}}, \
  and\ \bibinfo {author} {\bibfnamefont {E.~M.}\ \bibnamefont {Gauger}},\
  }\bibfield  {title} {\enquote {\bibinfo {title} {{Marcus Theory of
  Thermoelectricity in Molecular Junctions}},}\ }\href {\doibase
  10.1021/acs.jpcc.8b12163} {\bibfield  {journal} {\bibinfo  {journal} {J.
  Phys. Chem. C}\ }\textbf {\bibinfo {volume} {123}},\ \bibinfo {pages}
  {4103--4108} (\bibinfo {year} {2019})}\BibitemShut {NoStop}%
\bibitem [{\citenamefont {Kim}\ \emph {et~al.}(2014)\citenamefont {Kim},
  \citenamefont {Jeong}, \citenamefont {Kim}, \citenamefont {Lee},\ and\
  \citenamefont {Reddy}}]{Kim2014}%
  \BibitemOpen
  \bibfield  {author} {\bibinfo {author} {\bibfnamefont {Y.}~\bibnamefont
  {Kim}}, \bibinfo {author} {\bibfnamefont {W.}~\bibnamefont {Jeong}}, \bibinfo
  {author} {\bibfnamefont {K.}~\bibnamefont {Kim}}, \bibinfo {author}
  {\bibfnamefont {W.}~\bibnamefont {Lee}}, \ and\ \bibinfo {author}
  {\bibfnamefont {P.}~\bibnamefont {Reddy}},\ }\bibfield  {title} {\enquote
  {\bibinfo {title} {{Electrostatic control of thermoelectricity in molecular
  junctions}},}\ }\href {\doibase 10.1038/nnano.2014.209} {\bibfield  {journal}
  {\bibinfo  {journal} {Nat. Nanotechnol.}\ }\textbf {\bibinfo {volume} {9}},\
  \bibinfo {pages} {881--885} (\bibinfo {year} {2014})}\BibitemShut {NoStop}%
\bibitem [{\citenamefont {Dorsch}, \citenamefont {Fahlvik},\ and\ \citenamefont
  {Burke}(2021)}]{Dorsch2021}%
  \BibitemOpen
  \bibfield  {author} {\bibinfo {author} {\bibfnamefont {S.}~\bibnamefont
  {Dorsch}}, \bibinfo {author} {\bibfnamefont {S.}~\bibnamefont {Fahlvik}}, \
  and\ \bibinfo {author} {\bibfnamefont {A.}~\bibnamefont {Burke}},\ }\bibfield
   {title} {\enquote {\bibinfo {title} {{Characterization of electrostatically
  defined bottom-heated InAs nanowire quantum dot systems}},}\ }\href {\doibase
  10.1088/1367-2630/ac434c} {\bibfield  {journal} {\bibinfo  {journal} {New J.
  Phys.}\ }\textbf {\bibinfo {volume} {23}} (\bibinfo {year} {2021}),\
  10.1088/1367-2630/ac434c}\BibitemShut {NoStop}%
\bibitem [{\citenamefont {Kitaev}(2003)}]{Kitaev2003}%
  \BibitemOpen
  \bibfield  {author} {\bibinfo {author} {\bibfnamefont {A.}~\bibnamefont
  {Kitaev}},\ }\bibfield  {title} {\enquote {\bibinfo {title} {Fault-tolerant
  quantum computation by anyons},}\ }\href {\doibase
  https://doi.org/10.1016/S0003-4916(02)00018-0} {\bibfield  {journal}
  {\bibinfo  {journal} {Annals of Physics}\ }\textbf {\bibinfo {volume}
  {303}},\ \bibinfo {pages} {2--30} (\bibinfo {year} {2003})}\BibitemShut
  {NoStop}%
\bibitem [{\citenamefont {Lian}\ \emph {et~al.}(2018)\citenamefont {Lian},
  \citenamefont {Sun}, \citenamefont {Vaezi}, \citenamefont {Qi},\ and\
  \citenamefont {Zhang}}]{Lian2018}%
  \BibitemOpen
  \bibfield  {author} {\bibinfo {author} {\bibfnamefont {B.}~\bibnamefont
  {Lian}}, \bibinfo {author} {\bibfnamefont {X.-Q.}\ \bibnamefont {Sun}},
  \bibinfo {author} {\bibfnamefont {A.}~\bibnamefont {Vaezi}}, \bibinfo
  {author} {\bibfnamefont {X.-L.}\ \bibnamefont {Qi}}, \ and\ \bibinfo {author}
  {\bibfnamefont {S.-C.}\ \bibnamefont {Zhang}},\ }\bibfield  {title} {\enquote
  {\bibinfo {title} {Topological quantum computation based on chiral majorana
  fermions},}\ }\href {\doibase 10.1073/pnas.1810003115} {\bibfield  {journal}
  {\bibinfo  {journal} {Proceedings of the National Academy of Sciences}\
  }\textbf {\bibinfo {volume} {115}},\ \bibinfo {pages} {10938--10942}
  (\bibinfo {year} {2018})},\ \Eprint
  {http://arxiv.org/abs/https://www.pnas.org/doi/pdf/10.1073/pnas.1810003115}
  {https://www.pnas.org/doi/pdf/10.1073/pnas.1810003115} \BibitemShut {NoStop}%
\bibitem [{\citenamefont {Thorp}(2021)}]{Holden2021}%
  \BibitemOpen
  \bibfield  {author} {\bibinfo {author} {\bibfnamefont {H.~H.}\ \bibnamefont
  {Thorp}},\ }\bibfield  {title} {\enquote {\bibinfo {title} {Editorial
  expression of concern},}\ }\href {\doibase 10.1126/science.abn5849}
  {\bibfield  {journal} {\bibinfo  {journal} {Science}\ }\textbf {\bibinfo
  {volume} {374}},\ \bibinfo {pages} {1454--1454} (\bibinfo {year} {2021})},\
  \Eprint
  {http://arxiv.org/abs/https://www.science.org/doi/pdf/10.1126/science.abn5849}
  {https://www.science.org/doi/pdf/10.1126/science.abn5849} \BibitemShut
  {NoStop}%
\bibitem [{\citenamefont {Kayyalha}\ \emph {et~al.}(2020)\citenamefont
  {Kayyalha}, \citenamefont {Xiao}, \citenamefont {Zhang}, \citenamefont
  {Shin}, \citenamefont {Jiang}, \citenamefont {Wang}, \citenamefont {Zhao},
  \citenamefont {Xiao}, \citenamefont {Zhang}, \citenamefont {Fijalkowski},
  \citenamefont {Mandal}, \citenamefont {Winnerlein}, \citenamefont {Gould},
  \citenamefont {Li}, \citenamefont {Molenkamp}, \citenamefont {Chan},
  \citenamefont {Samarth},\ and\ \citenamefont {Chang}}]{Kayyalha2020}%
  \BibitemOpen
  \bibfield  {author} {\bibinfo {author} {\bibfnamefont {M.}~\bibnamefont
  {Kayyalha}}, \bibinfo {author} {\bibfnamefont {D.}~\bibnamefont {Xiao}},
  \bibinfo {author} {\bibfnamefont {R.}~\bibnamefont {Zhang}}, \bibinfo
  {author} {\bibfnamefont {J.}~\bibnamefont {Shin}}, \bibinfo {author}
  {\bibfnamefont {J.}~\bibnamefont {Jiang}}, \bibinfo {author} {\bibfnamefont
  {F.}~\bibnamefont {Wang}}, \bibinfo {author} {\bibfnamefont {Y.-F.}\
  \bibnamefont {Zhao}}, \bibinfo {author} {\bibfnamefont {R.}~\bibnamefont
  {Xiao}}, \bibinfo {author} {\bibfnamefont {L.}~\bibnamefont {Zhang}},
  \bibinfo {author} {\bibfnamefont {K.~M.}\ \bibnamefont {Fijalkowski}},
  \bibinfo {author} {\bibfnamefont {P.}~\bibnamefont {Mandal}}, \bibinfo
  {author} {\bibfnamefont {M.}~\bibnamefont {Winnerlein}}, \bibinfo {author}
  {\bibfnamefont {C.}~\bibnamefont {Gould}}, \bibinfo {author} {\bibfnamefont
  {Q.}~\bibnamefont {Li}}, \bibinfo {author} {\bibfnamefont {L.~W.}\
  \bibnamefont {Molenkamp}}, \bibinfo {author} {\bibfnamefont {M.~H.~W.}\
  \bibnamefont {Chan}}, \bibinfo {author} {\bibfnamefont {N.}~\bibnamefont
  {Samarth}}, \ and\ \bibinfo {author} {\bibfnamefont {C.-Z.}\ \bibnamefont
  {Chang}},\ }\bibfield  {title} {\enquote {\bibinfo {title} {Absence of
  evidence for chiral majorana modes in quantum anomalous hall-superconductor
  devices},}\ }\href {\doibase 10.1126/science.aax6361} {\bibfield  {journal}
  {\bibinfo  {journal} {Science}\ }\textbf {\bibinfo {volume} {367}},\ \bibinfo
  {pages} {64--67} (\bibinfo {year} {2020})},\ \Eprint
  {http://arxiv.org/abs/https://www.science.org/doi/pdf/10.1126/science.aax6361}
  {https://www.science.org/doi/pdf/10.1126/science.aax6361} \BibitemShut
  {NoStop}%
\bibitem [{\citenamefont {Zhang}\ \emph {et~al.}(2021)\citenamefont {Zhang},
  \citenamefont {Liu}, \citenamefont {Gazibegovic}, \citenamefont {Xu},
  \citenamefont {Logan}, \citenamefont {Wang}, \citenamefont {van Loo},
  \citenamefont {Bommer}, \citenamefont {de~Moor}, \citenamefont {Car},
  \citenamefont {Op~het Veld}, \citenamefont {van Veldhoven}, \citenamefont
  {Koelling}, \citenamefont {Verheijen}, \citenamefont {Pendharkar},
  \citenamefont {Pennachio}, \citenamefont {Shojaei}, \citenamefont {Lee},
  \citenamefont {Palmstr{\o}m}, \citenamefont {Bakkers}, \citenamefont
  {Das~Sarma},\ and\ \citenamefont {Kouwenhoven}}]{Zhang2021}%
  \BibitemOpen
  \bibfield  {author} {\bibinfo {author} {\bibfnamefont {H.}~\bibnamefont
  {Zhang}}, \bibinfo {author} {\bibfnamefont {C.-X.}\ \bibnamefont {Liu}},
  \bibinfo {author} {\bibfnamefont {S.}~\bibnamefont {Gazibegovic}}, \bibinfo
  {author} {\bibfnamefont {D.}~\bibnamefont {Xu}}, \bibinfo {author}
  {\bibfnamefont {J.~A.}\ \bibnamefont {Logan}}, \bibinfo {author}
  {\bibfnamefont {G.}~\bibnamefont {Wang}}, \bibinfo {author} {\bibfnamefont
  {N.}~\bibnamefont {van Loo}}, \bibinfo {author} {\bibfnamefont {J.~D.~S.}\
  \bibnamefont {Bommer}}, \bibinfo {author} {\bibfnamefont {M.~W.~A.}\
  \bibnamefont {de~Moor}}, \bibinfo {author} {\bibfnamefont {D.}~\bibnamefont
  {Car}}, \bibinfo {author} {\bibfnamefont {R.~L.~M.}\ \bibnamefont {Op~het
  Veld}}, \bibinfo {author} {\bibfnamefont {P.~J.}\ \bibnamefont {van
  Veldhoven}}, \bibinfo {author} {\bibfnamefont {S.}~\bibnamefont {Koelling}},
  \bibinfo {author} {\bibfnamefont {M.~A.}\ \bibnamefont {Verheijen}}, \bibinfo
  {author} {\bibfnamefont {M.}~\bibnamefont {Pendharkar}}, \bibinfo {author}
  {\bibfnamefont {D.~J.}\ \bibnamefont {Pennachio}}, \bibinfo {author}
  {\bibfnamefont {B.}~\bibnamefont {Shojaei}}, \bibinfo {author} {\bibfnamefont
  {J.~S.}\ \bibnamefont {Lee}}, \bibinfo {author} {\bibfnamefont {C.~J.}\
  \bibnamefont {Palmstr{\o}m}}, \bibinfo {author} {\bibfnamefont {E.~P. A.~M.}\
  \bibnamefont {Bakkers}}, \bibinfo {author} {\bibfnamefont {S.}~\bibnamefont
  {Das~Sarma}}, \ and\ \bibinfo {author} {\bibfnamefont {L.~P.}\ \bibnamefont
  {Kouwenhoven}},\ }\bibfield  {title} {\enquote {\bibinfo {title} {Retraction
  note: Quantized majorana conductance},}\ }\href {\doibase
  10.1038/s41586-021-03373-x} {\bibfield  {journal} {\bibinfo  {journal}
  {Nature}\ }\textbf {\bibinfo {volume} {591}},\ \bibinfo {pages} {E30--E30}
  (\bibinfo {year} {2021})}\BibitemShut {NoStop}%
\bibitem [{\citenamefont {Sills}\ and\ \citenamefont
  {Thorp}(2021)}]{Sills2021}%
  \BibitemOpen
  \bibfield  {author} {\bibinfo {author} {\bibfnamefont {J.}~\bibnamefont
  {Sills}}\ and\ \bibinfo {author} {\bibfnamefont {H.~H.}\ \bibnamefont
  {Thorp}},\ }\bibfield  {title} {\enquote {\bibinfo {title} {Editorial
  expression of concern},}\ }\href {\doibase 10.1126/science.abl5286}
  {\bibfield  {journal} {\bibinfo  {journal} {Science}\ }\textbf {\bibinfo
  {volume} {373}},\ \bibinfo {pages} {500--500} (\bibinfo {year} {2021})},\
  \Eprint
  {http://arxiv.org/abs/https://www.science.org/doi/pdf/10.1126/science.abl5286}
  {https://www.science.org/doi/pdf/10.1126/science.abl5286} \BibitemShut
  {NoStop}%
\bibitem [{\citenamefont {Ben-Shach}\ \emph {et~al.}(2013)\citenamefont
  {Ben-Shach}, \citenamefont {Laumann}, \citenamefont {Neder}, \citenamefont
  {Yacoby},\ and\ \citenamefont {Halperin}}]{BenShach2013}%
  \BibitemOpen
  \bibfield  {author} {\bibinfo {author} {\bibfnamefont {G.}~\bibnamefont
  {Ben-Shach}}, \bibinfo {author} {\bibfnamefont {C.~R.}\ \bibnamefont
  {Laumann}}, \bibinfo {author} {\bibfnamefont {I.}~\bibnamefont {Neder}},
  \bibinfo {author} {\bibfnamefont {A.}~\bibnamefont {Yacoby}}, \ and\ \bibinfo
  {author} {\bibfnamefont {B.~I.}\ \bibnamefont {Halperin}},\ }\bibfield
  {title} {\enquote {\bibinfo {title} {Detecting non-abelian anyons by charging
  spectroscopy},}\ }\href {\doibase 10.1103/PhysRevLett.110.106805} {\bibfield
  {journal} {\bibinfo  {journal} {Phys. Rev. Lett.}\ }\textbf {\bibinfo
  {volume} {110}},\ \bibinfo {pages} {106805} (\bibinfo {year}
  {2013})}\BibitemShut {NoStop}%
\bibitem [{\citenamefont {Smirnov}(2015)}]{Smirnov2015}%
  \BibitemOpen
  \bibfield  {author} {\bibinfo {author} {\bibfnamefont {S.}~\bibnamefont
  {Smirnov}},\ }\bibfield  {title} {\enquote {\bibinfo {title} {Majorana
  tunneling entropy},}\ }\href {\doibase 10.1103/PhysRevB.92.195312} {\bibfield
   {journal} {\bibinfo  {journal} {Phys. Rev. B}\ }\textbf {\bibinfo {volume}
  {92}},\ \bibinfo {pages} {195312} (\bibinfo {year} {2015})}\BibitemShut
  {NoStop}%
\bibitem [{\citenamefont {Smirnov}(2021{\natexlab{a}})}]{Smirnov2021a}%
  \BibitemOpen
  \bibfield  {author} {\bibinfo {author} {\bibfnamefont {S.}~\bibnamefont
  {Smirnov}},\ }\bibfield  {title} {\enquote {\bibinfo {title} {Majorana
  entropy revival via tunneling phases},}\ }\href {\doibase
  10.1103/PhysRevB.103.075440} {\bibfield  {journal} {\bibinfo  {journal}
  {Phys. Rev. B}\ }\textbf {\bibinfo {volume} {103}},\ \bibinfo {pages}
  {075440} (\bibinfo {year} {2021}{\natexlab{a}})}\BibitemShut {NoStop}%
\bibitem [{\citenamefont {Gau}\ \emph {et~al.}(2020{\natexlab{a}})\citenamefont
  {Gau}, \citenamefont {Egger}, \citenamefont {Zazunov},\ and\ \citenamefont
  {Gefen}}]{Gau2020a}%
  \BibitemOpen
  \bibfield  {author} {\bibinfo {author} {\bibfnamefont {M.}~\bibnamefont
  {Gau}}, \bibinfo {author} {\bibfnamefont {R.}~\bibnamefont {Egger}}, \bibinfo
  {author} {\bibfnamefont {A.}~\bibnamefont {Zazunov}}, \ and\ \bibinfo
  {author} {\bibfnamefont {Y.}~\bibnamefont {Gefen}},\ }\bibfield  {title}
  {\enquote {\bibinfo {title} {Towards dark space stabilization and
  manipulation in driven dissipative majorana platforms},}\ }\href {\doibase
  10.1103/PhysRevB.102.134501} {\bibfield  {journal} {\bibinfo  {journal}
  {Phys. Rev. B}\ }\textbf {\bibinfo {volume} {102}},\ \bibinfo {pages}
  {134501} (\bibinfo {year} {2020}{\natexlab{a}})}\BibitemShut {NoStop}%
\bibitem [{\citenamefont {Gau}\ \emph {et~al.}(2020{\natexlab{b}})\citenamefont
  {Gau}, \citenamefont {Egger}, \citenamefont {Zazunov},\ and\ \citenamefont
  {Gefen}}]{Gau2020b}%
  \BibitemOpen
  \bibfield  {author} {\bibinfo {author} {\bibfnamefont {M.}~\bibnamefont
  {Gau}}, \bibinfo {author} {\bibfnamefont {R.}~\bibnamefont {Egger}}, \bibinfo
  {author} {\bibfnamefont {A.}~\bibnamefont {Zazunov}}, \ and\ \bibinfo
  {author} {\bibfnamefont {Y.}~\bibnamefont {Gefen}},\ }\bibfield  {title}
  {\enquote {\bibinfo {title} {Driven dissipative majorana dark spaces},}\
  }\href {\doibase 10.1103/PhysRevLett.125.147701} {\bibfield  {journal}
  {\bibinfo  {journal} {Phys. Rev. Lett.}\ }\textbf {\bibinfo {volume} {125}},\
  \bibinfo {pages} {147701} (\bibinfo {year} {2020}{\natexlab{b}})}\BibitemShut
  {NoStop}%
\bibitem [{\citenamefont {Smirnov}(2021{\natexlab{b}})}]{Smirnov2021b}%
  \BibitemOpen
  \bibfield  {author} {\bibinfo {author} {\bibfnamefont {S.}~\bibnamefont
  {Smirnov}},\ }\bibfield  {title} {\enquote {\bibinfo {title} {Majorana
  ensembles with fractional entropy and conductance in nanoscopic systems},}\
  }\href {\doibase 10.1103/PhysRevB.104.205406} {\bibfield  {journal} {\bibinfo
   {journal} {Phys. Rev. B}\ }\textbf {\bibinfo {volume} {104}},\ \bibinfo
  {pages} {205406} (\bibinfo {year} {2021}{\natexlab{b}})}\BibitemShut
  {NoStop}%
\bibitem [{\citenamefont {Hsu}\ \emph {et~al.}(2022)\citenamefont {Hsu},
  \citenamefont {Costi}, \citenamefont {Vogel}, \citenamefont {Wegeberg},
  \citenamefont {Mayor}, \citenamefont {van~der Zant},\ and\ \citenamefont
  {Gehring}}]{Hsu2022}%
  \BibitemOpen
  \bibfield  {author} {\bibinfo {author} {\bibfnamefont {C.}~\bibnamefont
  {Hsu}}, \bibinfo {author} {\bibfnamefont {T.~A.}\ \bibnamefont {Costi}},
  \bibinfo {author} {\bibfnamefont {D.}~\bibnamefont {Vogel}}, \bibinfo
  {author} {\bibfnamefont {C.}~\bibnamefont {Wegeberg}}, \bibinfo {author}
  {\bibfnamefont {M.}~\bibnamefont {Mayor}}, \bibinfo {author} {\bibfnamefont
  {H.~S.~J.}\ \bibnamefont {van~der Zant}}, \ and\ \bibinfo {author}
  {\bibfnamefont {P.}~\bibnamefont {Gehring}},\ }\bibfield  {title} {\enquote
  {\bibinfo {title} {Magnetic-field universality of the kondo effect revealed
  by thermocurrent spectroscopy},}\ }\href {\doibase
  10.1103/PhysRevLett.128.147701} {\bibfield  {journal} {\bibinfo  {journal}
  {Phys. Rev. Lett.}\ }\textbf {\bibinfo {volume} {128}},\ \bibinfo {pages}
  {147701} (\bibinfo {year} {2022})}\BibitemShut {NoStop}%
\bibitem [{\citenamefont {Iftikhar}\ \emph {et~al.}(2015)\citenamefont
  {Iftikhar}, \citenamefont {Jezouin}, \citenamefont {Anthore}, \citenamefont
  {Gennser}, \citenamefont {Parmentier}, \citenamefont {Cavanna},\ and\
  \citenamefont {Pierre}}]{Iftikhar2015}%
  \BibitemOpen
  \bibfield  {author} {\bibinfo {author} {\bibfnamefont {Z.}~\bibnamefont
  {Iftikhar}}, \bibinfo {author} {\bibfnamefont {S.}~\bibnamefont {Jezouin}},
  \bibinfo {author} {\bibfnamefont {A.}~\bibnamefont {Anthore}}, \bibinfo
  {author} {\bibfnamefont {U.}~\bibnamefont {Gennser}}, \bibinfo {author}
  {\bibfnamefont {F.~D.}\ \bibnamefont {Parmentier}}, \bibinfo {author}
  {\bibfnamefont {A.}~\bibnamefont {Cavanna}}, \ and\ \bibinfo {author}
  {\bibfnamefont {F.}~\bibnamefont {Pierre}},\ }\bibfield  {title} {\enquote
  {\bibinfo {title} {Two-channel kondo effect and renormalization flow with
  macroscopic quantum charge states},}\ }\href {\doibase 10.1038/nature15384}
  {\bibfield  {journal} {\bibinfo  {journal} {Nature}\ }\textbf {\bibinfo
  {volume} {526}},\ \bibinfo {pages} {233--236} (\bibinfo {year}
  {2015})}\BibitemShut {NoStop}%
\bibitem [{\citenamefont {Iftikhar}\ \emph {et~al.}(2018)\citenamefont
  {Iftikhar}, \citenamefont {Anthore}, \citenamefont {Mitchell}, \citenamefont
  {Parmentier}, \citenamefont {Gennser}, \citenamefont {Ouerghi}, \citenamefont
  {Cavanna}, \citenamefont {Mora}, \citenamefont {Simon},\ and\ \citenamefont
  {Pierre}}]{Iftikhar2018}%
  \BibitemOpen
  \bibfield  {author} {\bibinfo {author} {\bibfnamefont {Z.}~\bibnamefont
  {Iftikhar}}, \bibinfo {author} {\bibfnamefont {A.}~\bibnamefont {Anthore}},
  \bibinfo {author} {\bibfnamefont {A.~K.}\ \bibnamefont {Mitchell}}, \bibinfo
  {author} {\bibfnamefont {F.~D.}\ \bibnamefont {Parmentier}}, \bibinfo
  {author} {\bibfnamefont {U.}~\bibnamefont {Gennser}}, \bibinfo {author}
  {\bibfnamefont {A.}~\bibnamefont {Ouerghi}}, \bibinfo {author} {\bibfnamefont
  {A.}~\bibnamefont {Cavanna}}, \bibinfo {author} {\bibfnamefont
  {C.}~\bibnamefont {Mora}}, \bibinfo {author} {\bibfnamefont {P.}~\bibnamefont
  {Simon}}, \ and\ \bibinfo {author} {\bibfnamefont {F.}~\bibnamefont
  {Pierre}},\ }\bibfield  {title} {\enquote {\bibinfo {title} {Tunable quantum
  criticality and super-ballistic transport in a ``charge'' kondo circuit},}\
  }\href {\doibase 10.1126/science.aan5592} {\bibfield  {journal} {\bibinfo
  {journal} {Science}\ }\textbf {\bibinfo {volume} {360}},\ \bibinfo {pages}
  {1315--1320} (\bibinfo {year} {2018})},\ \Eprint
  {http://arxiv.org/abs/https://www.science.org/doi/pdf/10.1126/science.aan5592}
  {https://www.science.org/doi/pdf/10.1126/science.aan5592} \BibitemShut
  {NoStop}%
\bibitem [{\citenamefont {Meirav}, \citenamefont {Kastner},\ and\ \citenamefont
  {Wind}(1990)}]{Meirav90}%
  \BibitemOpen
  \bibfield  {author} {\bibinfo {author} {\bibfnamefont {U.}~\bibnamefont
  {Meirav}}, \bibinfo {author} {\bibfnamefont {M.~A.}\ \bibnamefont {Kastner}},
  \ and\ \bibinfo {author} {\bibfnamefont {S.~J.}\ \bibnamefont {Wind}},\
  }\bibfield  {title} {\enquote {\bibinfo {title} {Single-electron charging and
  periodic conductance resonances in gaas nanostructures},}\ }\href {\doibase
  10.1103/PhysRevLett.65.771} {\bibfield  {journal} {\bibinfo  {journal} {Phys.
  Rev. Lett.}\ }\textbf {\bibinfo {volume} {65}},\ \bibinfo {pages} {771--774}
  (\bibinfo {year} {1990})}\BibitemShut {NoStop}%
\bibitem [{\citenamefont {Reed}\ \emph {et~al.}(1988)\citenamefont {Reed},
  \citenamefont {Randall}, \citenamefont {Aggarwal}, \citenamefont {Matyi},
  \citenamefont {Moore},\ and\ \citenamefont {Wetsel}}]{Reed88}%
  \BibitemOpen
  \bibfield  {author} {\bibinfo {author} {\bibfnamefont {M.~A.}\ \bibnamefont
  {Reed}}, \bibinfo {author} {\bibfnamefont {J.~N.}\ \bibnamefont {Randall}},
  \bibinfo {author} {\bibfnamefont {R.~J.}\ \bibnamefont {Aggarwal}}, \bibinfo
  {author} {\bibfnamefont {R.~J.}\ \bibnamefont {Matyi}}, \bibinfo {author}
  {\bibfnamefont {T.~M.}\ \bibnamefont {Moore}}, \ and\ \bibinfo {author}
  {\bibfnamefont {A.~E.}\ \bibnamefont {Wetsel}},\ }\bibfield  {title}
  {\enquote {\bibinfo {title} {Observation of discrete electronic states in a
  zero-dimensional semiconductor nanostructure},}\ }\href {\doibase
  10.1103/PhysRevLett.60.535} {\bibfield  {journal} {\bibinfo  {journal} {Phys.
  Rev. Lett.}\ }\textbf {\bibinfo {volume} {60}},\ \bibinfo {pages} {535--537}
  (\bibinfo {year} {1988})}\BibitemShut {NoStop}%
\bibitem [{\citenamefont {Thelander}\ \emph {et~al.}(2003)\citenamefont
  {Thelander}, \citenamefont {Mårtensson}, \citenamefont {Björk},
  \citenamefont {Ohlsson}, \citenamefont {Larsson}, \citenamefont
  {Wallenberg},\ and\ \citenamefont {Samuelson}}]{Thelander03}%
  \BibitemOpen
  \bibfield  {author} {\bibinfo {author} {\bibfnamefont {C.}~\bibnamefont
  {Thelander}}, \bibinfo {author} {\bibfnamefont {T.}~\bibnamefont
  {Mårtensson}}, \bibinfo {author} {\bibfnamefont {M.~T.}\ \bibnamefont
  {Björk}}, \bibinfo {author} {\bibfnamefont {B.~J.}\ \bibnamefont {Ohlsson}},
  \bibinfo {author} {\bibfnamefont {M.~W.}\ \bibnamefont {Larsson}}, \bibinfo
  {author} {\bibfnamefont {L.~R.}\ \bibnamefont {Wallenberg}}, \ and\ \bibinfo
  {author} {\bibfnamefont {L.}~\bibnamefont {Samuelson}},\ }\bibfield  {title}
  {\enquote {\bibinfo {title} {Single-electron transistors in heterostructure
  nanowires},}\ }\href {\doibase 10.1063/1.1606889} {\bibfield  {journal}
  {\bibinfo  {journal} {Applied Physics Letters}\ }\textbf {\bibinfo {volume}
  {83}},\ \bibinfo {pages} {2052--2054} (\bibinfo {year} {2003})},\ \Eprint
  {http://arxiv.org/abs/https://doi.org/10.1063/1.1606889}
  {https://doi.org/10.1063/1.1606889} \BibitemShut {NoStop}%
\bibitem [{\citenamefont {Liang}\ \emph {et~al.}(2002)\citenamefont {Liang},
  \citenamefont {Shores}, \citenamefont {Bockrath}, \citenamefont {Long},\ and\
  \citenamefont {Park}}]{Liang2002}%
  \BibitemOpen
  \bibfield  {author} {\bibinfo {author} {\bibfnamefont {W.}~\bibnamefont
  {Liang}}, \bibinfo {author} {\bibfnamefont {M.~P.}\ \bibnamefont {Shores}},
  \bibinfo {author} {\bibfnamefont {M.}~\bibnamefont {Bockrath}}, \bibinfo
  {author} {\bibfnamefont {J.~R.}\ \bibnamefont {Long}}, \ and\ \bibinfo
  {author} {\bibfnamefont {H.}~\bibnamefont {Park}},\ }\bibfield  {title}
  {\enquote {\bibinfo {title} {Kondo resonance in a single-molecule
  transistor},}\ }\href {\doibase 10.1038/nature00790} {\bibfield  {journal}
  {\bibinfo  {journal} {Nature}\ }\textbf {\bibinfo {volume} {417}},\ \bibinfo
  {pages} {725--729} (\bibinfo {year} {2002})}\BibitemShut {NoStop}%
\bibitem [{\citenamefont {Gehring}, \citenamefont {Thijssen},\ and\
  \citenamefont {van~der Zant}(2019)}]{Gehring2019}%
  \BibitemOpen
  \bibfield  {author} {\bibinfo {author} {\bibfnamefont {P.}~\bibnamefont
  {Gehring}}, \bibinfo {author} {\bibfnamefont {J.~M.}\ \bibnamefont
  {Thijssen}}, \ and\ \bibinfo {author} {\bibfnamefont {H.~S.~J.}\ \bibnamefont
  {van~der Zant}},\ }\bibfield  {title} {\enquote {\bibinfo {title}
  {Single-molecule quantum-transport phenomena in break junctions},}\ }\href
  {\doibase 10.1038/s42254-019-0055-1} {\bibfield  {journal} {\bibinfo
  {journal} {Nature Reviews Physics}\ }\textbf {\bibinfo {volume} {1}},\
  \bibinfo {pages} {381--396} (\bibinfo {year} {2019})}\BibitemShut {NoStop}%
\bibitem [{\citenamefont {Park}, \citenamefont {Kang},\ and\ \citenamefont
  {Yoon}(2019)}]{Par19}%
  \BibitemOpen
  \bibfield  {author} {\bibinfo {author} {\bibfnamefont {S.}~\bibnamefont
  {Park}}, \bibinfo {author} {\bibfnamefont {S.}~\bibnamefont {Kang}}, \ and\
  \bibinfo {author} {\bibfnamefont {H.~J.}\ \bibnamefont {Yoon}},\ }\bibfield
  {title} {\enquote {\bibinfo {title} {Power factor of one molecule thick films
  and length dependence},}\ }\href {\doibase 10.1021/acscentsci.9b01042}
  {\bibfield  {journal} {\bibinfo  {journal} {ACS Central Science}\ }\textbf
  {\bibinfo {volume} {5}},\ \bibinfo {pages} {1975--1982} (\bibinfo {year}
  {2019})},\ \bibinfo {note} {pMID: 31893227},\ \Eprint
  {http://arxiv.org/abs/https://doi.org/10.1021/acscentsci.9b01042}
  {https://doi.org/10.1021/acscentsci.9b01042} \BibitemShut {NoStop}%
\bibitem [{\citenamefont {Widawsky}\ \emph {et~al.}(2012)\citenamefont
  {Widawsky}, \citenamefont {Darancet}, \citenamefont {Neaton},\ and\
  \citenamefont {Venkataraman}}]{Wid12}%
  \BibitemOpen
  \bibfield  {author} {\bibinfo {author} {\bibfnamefont {J.~R.}\ \bibnamefont
  {Widawsky}}, \bibinfo {author} {\bibfnamefont {P.}~\bibnamefont {Darancet}},
  \bibinfo {author} {\bibfnamefont {J.~B.}\ \bibnamefont {Neaton}}, \ and\
  \bibinfo {author} {\bibfnamefont {L.}~\bibnamefont {Venkataraman}},\
  }\bibfield  {title} {\enquote {\bibinfo {title} {Simultaneous determination
  of conductance and thermopower of single molecule junctions},}\ }\href
  {\doibase 10.1021/nl203634m} {\bibfield  {journal} {\bibinfo  {journal} {Nano
  Letters}\ }\textbf {\bibinfo {volume} {12}},\ \bibinfo {pages} {354--358}
  (\bibinfo {year} {2012})},\ \bibinfo {note} {pMID: 22128800},\ \Eprint
  {http://arxiv.org/abs/https://doi.org/10.1021/nl203634m}
  {https://doi.org/10.1021/nl203634m} \BibitemShut {NoStop}%
\bibitem [{\citenamefont {Lambert}, \citenamefont {Sadeghi},\ and\
  \citenamefont {Al-Galiby}(2016)}]{Lam16}%
  \BibitemOpen
  \bibfield  {author} {\bibinfo {author} {\bibfnamefont {C.~J.}\ \bibnamefont
  {Lambert}}, \bibinfo {author} {\bibfnamefont {H.}~\bibnamefont {Sadeghi}}, \
  and\ \bibinfo {author} {\bibfnamefont {Q.~H.}\ \bibnamefont {Al-Galiby}},\
  }\bibfield  {title} {\enquote {\bibinfo {title}
  {Quantum-interference-enhanced thermoelectricity in single molecules and
  molecular films},}\ }\href {\doibase
  https://doi.org/10.1016/j.crhy.2016.08.003} {\bibfield  {journal} {\bibinfo
  {journal} {Comptes Rendus Physique}\ }\textbf {\bibinfo {volume} {17}},\
  \bibinfo {pages} {1084--1095} (\bibinfo {year} {2016})},\ \bibinfo {note}
  {mesoscopic thermoelectric phenomena / Phénomènes thermoélectriques
  mésoscopiques}\BibitemShut {NoStop}%
\bibitem [{\citenamefont {Reddy}\ \emph {et~al.}(2007)\citenamefont {Reddy},
  \citenamefont {Jang}, \citenamefont {Segalman},\ and\ \citenamefont
  {Majumdar}}]{Reddy2007}%
  \BibitemOpen
  \bibfield  {author} {\bibinfo {author} {\bibfnamefont {P.}~\bibnamefont
  {Reddy}}, \bibinfo {author} {\bibfnamefont {S.-Y.}\ \bibnamefont {Jang}},
  \bibinfo {author} {\bibfnamefont {R.~A.}\ \bibnamefont {Segalman}}, \ and\
  \bibinfo {author} {\bibfnamefont {A.}~\bibnamefont {Majumdar}},\ }\bibfield
  {title} {\enquote {\bibinfo {title} {Thermoelectricity in molecular
  junctions},}\ }\href {\doibase 10.1126/science.1137149} {\bibfield  {journal}
  {\bibinfo  {journal} {Science}\ }\textbf {\bibinfo {volume} {315}},\ \bibinfo
  {pages} {1568--1571} (\bibinfo {year} {2007})},\ \Eprint
  {http://arxiv.org/abs/https://www.science.org/doi/pdf/10.1126/science.1137149}
  {https://www.science.org/doi/pdf/10.1126/science.1137149} \BibitemShut
  {NoStop}%
\bibitem [{\citenamefont {Rincón-García}\ \emph {et~al.}(2016)\citenamefont
  {Rincón-García}, \citenamefont {Evangeli}, \citenamefont
  {Rubio-Bollinger},\ and\ \citenamefont {Agraït}}]{Rincon2016}%
  \BibitemOpen
  \bibfield  {author} {\bibinfo {author} {\bibfnamefont {L.}~\bibnamefont
  {Rincón-García}}, \bibinfo {author} {\bibfnamefont {C.}~\bibnamefont
  {Evangeli}}, \bibinfo {author} {\bibfnamefont {G.}~\bibnamefont
  {Rubio-Bollinger}}, \ and\ \bibinfo {author} {\bibfnamefont {N.}~\bibnamefont
  {Agraït}},\ }\bibfield  {title} {\enquote {\bibinfo {title} {Thermopower
  measurements in molecular junctions},}\ }\href {\doibase 10.1039/C6CS00141F}
  {\bibfield  {journal} {\bibinfo  {journal} {Chem. Soc. Rev.}\ }\textbf
  {\bibinfo {volume} {45}},\ \bibinfo {pages} {4285--4306} (\bibinfo {year}
  {2016})}\BibitemShut {NoStop}%
\bibitem [{\citenamefont {Cui}\ \emph {et~al.}(2017)\citenamefont {Cui},
  \citenamefont {Miao}, \citenamefont {Jiang}, \citenamefont {Meyhofer},\ and\
  \citenamefont {Reddy}}]{Cui2017}%
  \BibitemOpen
  \bibfield  {author} {\bibinfo {author} {\bibfnamefont {L.}~\bibnamefont
  {Cui}}, \bibinfo {author} {\bibfnamefont {R.}~\bibnamefont {Miao}}, \bibinfo
  {author} {\bibfnamefont {C.}~\bibnamefont {Jiang}}, \bibinfo {author}
  {\bibfnamefont {E.}~\bibnamefont {Meyhofer}}, \ and\ \bibinfo {author}
  {\bibfnamefont {P.}~\bibnamefont {Reddy}},\ }\bibfield  {title} {\enquote
  {\bibinfo {title} {Perspective: Thermal and thermoelectric transport in
  molecular junctions},}\ }\href {\doibase 10.1063/1.4976982} {\bibfield
  {journal} {\bibinfo  {journal} {The Journal of Chemical Physics}\ }\textbf
  {\bibinfo {volume} {146}},\ \bibinfo {pages} {092201} (\bibinfo {year}
  {2017})},\ \Eprint {http://arxiv.org/abs/https://doi.org/10.1063/1.4976982}
  {https://doi.org/10.1063/1.4976982} \BibitemShut {NoStop}%
\bibitem [{\citenamefont {Gehring}\ \emph {et~al.}(2017)\citenamefont
  {Gehring}, \citenamefont {Harzheim}, \citenamefont {Spièce}, \citenamefont
  {Sheng}, \citenamefont {Rogers}, \citenamefont {Evangeli}, \citenamefont
  {Mishra}, \citenamefont {Robinson}, \citenamefont {Porfyrakis}, \citenamefont
  {Warner}, \citenamefont {Kolosov}, \citenamefont {Briggs},\ and\
  \citenamefont {Mol}}]{Gehring2017}%
  \BibitemOpen
  \bibfield  {author} {\bibinfo {author} {\bibfnamefont {P.}~\bibnamefont
  {Gehring}}, \bibinfo {author} {\bibfnamefont {A.}~\bibnamefont {Harzheim}},
  \bibinfo {author} {\bibfnamefont {J.}~\bibnamefont {Spièce}}, \bibinfo
  {author} {\bibfnamefont {Y.}~\bibnamefont {Sheng}}, \bibinfo {author}
  {\bibfnamefont {G.}~\bibnamefont {Rogers}}, \bibinfo {author} {\bibfnamefont
  {C.}~\bibnamefont {Evangeli}}, \bibinfo {author} {\bibfnamefont
  {A.}~\bibnamefont {Mishra}}, \bibinfo {author} {\bibfnamefont {B.~J.}\
  \bibnamefont {Robinson}}, \bibinfo {author} {\bibfnamefont {K.}~\bibnamefont
  {Porfyrakis}}, \bibinfo {author} {\bibfnamefont {J.~H.}\ \bibnamefont
  {Warner}}, \bibinfo {author} {\bibfnamefont {O.~V.}\ \bibnamefont {Kolosov}},
  \bibinfo {author} {\bibfnamefont {G.~A.~D.}\ \bibnamefont {Briggs}}, \ and\
  \bibinfo {author} {\bibfnamefont {J.~A.}\ \bibnamefont {Mol}},\ }\bibfield
  {title} {\enquote {\bibinfo {title} {Field-effect control of
  graphene–fullerene thermoelectric nanodevices},}\ }\href {\doibase
  10.1021/acs.nanolett.7b03736} {\bibfield  {journal} {\bibinfo  {journal}
  {Nano Letters}\ }\textbf {\bibinfo {volume} {17}},\ \bibinfo {pages}
  {7055--7061} (\bibinfo {year} {2017})},\ \bibinfo {note} {pMID: 28982009},\
  \Eprint {http://arxiv.org/abs/https://doi.org/10.1021/acs.nanolett.7b03736}
  {https://doi.org/10.1021/acs.nanolett.7b03736} \BibitemShut {NoStop}%
\bibitem [{\citenamefont {Wang}, \citenamefont {Meyhofer},\ and\ \citenamefont
  {Reddy}(2020)}]{Wang2020}%
  \BibitemOpen
  \bibfield  {author} {\bibinfo {author} {\bibfnamefont {K.}~\bibnamefont
  {Wang}}, \bibinfo {author} {\bibfnamefont {E.}~\bibnamefont {Meyhofer}}, \
  and\ \bibinfo {author} {\bibfnamefont {P.}~\bibnamefont {Reddy}},\ }\bibfield
   {title} {\enquote {\bibinfo {title} {Thermal and thermoelectric properties
  of molecular junctions},}\ }\href {\doibase
  https://doi.org/10.1002/adfm.201904534} {\bibfield  {journal} {\bibinfo
  {journal} {Advanced Functional Materials}\ }\textbf {\bibinfo {volume}
  {30}},\ \bibinfo {pages} {1904534} (\bibinfo {year} {2020})},\ \Eprint
  {http://arxiv.org/abs/https://onlinelibrary.wiley.com/doi/pdf/10.1002/adfm.201904534}
  {https://onlinelibrary.wiley.com/doi/pdf/10.1002/adfm.201904534} \BibitemShut
  {NoStop}%
\bibitem [{\citenamefont {Micadei}\ \emph {et~al.}(2019)\citenamefont
  {Micadei}, \citenamefont {Peterson}, \citenamefont {Souza}, \citenamefont
  {Sarthour}, \citenamefont {Oliveira}, \citenamefont {Landi}, \citenamefont
  {Batalh{\~{a}}o}, \citenamefont {Serra},\ and\ \citenamefont
  {Lutz}}]{Micadei2017}%
  \BibitemOpen
  \bibfield  {author} {\bibinfo {author} {\bibfnamefont {K.}~\bibnamefont
  {Micadei}}, \bibinfo {author} {\bibfnamefont {J.~P.~S.}\ \bibnamefont
  {Peterson}}, \bibinfo {author} {\bibfnamefont {A.~M.}\ \bibnamefont {Souza}},
  \bibinfo {author} {\bibfnamefont {R.~S.}\ \bibnamefont {Sarthour}}, \bibinfo
  {author} {\bibfnamefont {I.~S.}\ \bibnamefont {Oliveira}}, \bibinfo {author}
  {\bibfnamefont {G.~T.}\ \bibnamefont {Landi}}, \bibinfo {author}
  {\bibfnamefont {T.~B.}\ \bibnamefont {Batalh{\~{a}}o}}, \bibinfo {author}
  {\bibfnamefont {R.~M.}\ \bibnamefont {Serra}}, \ and\ \bibinfo {author}
  {\bibfnamefont {E.}~\bibnamefont {Lutz}},\ }\bibfield  {title} {\enquote
  {\bibinfo {title} {{Reversing the direction of heat flow using quantum
  correlations}},}\ }\href {\doibase 10.1038/s41467-019-10333-7} {\bibfield
  {journal} {\bibinfo  {journal} {Nat. Commun.}\ }\textbf {\bibinfo {volume}
  {10}},\ \bibinfo {pages} {2456} (\bibinfo {year} {2019})},\ \Eprint
  {http://arxiv.org/abs/1711.03323v1} {arXiv:1711.03323v1} \BibitemShut
  {NoStop}%
\end{thebibliography}

%

\end{document}